\documentclass[12pt,twoside]{article}
\usepackage{fleqn,espcrc1} 
\usepackage{psfig}

\title{Low energy kaon photoproduction from nuclei\thanks{Supported 
by Forschungszentrum J\"ulich and DFG.}}

\author{A. Sibirtsev, W. Cassing, U. Mosel \\
\vspace{3mm} Institut f\"ur Theoretische Physik, 
Universit\"at Giessen \\ D-35392 Giessen, Germany}
\begin{document}
\maketitle

\begin{abstract}
We study   $K^+$-meson production in $\gamma{A}$ interaction at 
energies below the reaction threshold in free space. 
The Thomas-Fermi and spectral function approaches are used
for the calculations of the production process.
It is found that the measurement of the differential spectra may 
allow to reconstruct the production mechanism and to investigate 
the dispersion relations entering the production vertex.
It is shown that the contribution from secondary pion induced
reactions to the total kaon photoproduction is negligible for
$E_\gamma{<}$1.2~GeV so that strangeness production at low
energies is sensitive to the nuclear spectral function.
\end{abstract}

\vspace{4mm}


\section{Introduction}
The description of particle production in hadron-nucleus
collision can be considered relatively to an energy scale 
given by the binding energy of the target~\cite{Newton}. 
The impulse approximation is an appropriate basis
when the projectile energy is much higher than the  
binding energy and the target nucleons can be taken 
as  free particles. 
However, at low energies the nuclear structure
plays an essential role in the production process.
When the bombarding energies are below the reaction threshold
in free space the production process proceeds according to the
energy available from the target. Now the reaction mechanism
can no  longer be considered as  an interaction of the 
projectile with a quasi-free nucleon, but instead as an  
interaction with the whole nucleus, which might
be approximated by the sum over the interactions with one 
nucleon (single scattering approximation), two nucleons 
(two-nucleon current) and further many-body 
collisions~\cite{Newton,Kerman,Jackson}.  

In momentum space these many-body effects can be interpreted in 
terms of high momentum components of the nuclear wave function.
As found in Ref.~\cite{Atti} these high momentum components of 
the nuclear wave function can be reasonably well described by
short range two-nucleon correlations. In this sense the 
contribution from the high momentum components to  particle
production can be considered as a reaction mechanism due to the
two-nucleon current.

Here we study the $K^+$-meson  photoproduction from nuclei 
at energies below the $\gamma{p}\to K^+\Lambda$ reaction
threshold in free space. This reaction has a set of advantages 
for an investigation of the production mechanism. First, the
elementary production amplitude is reasonably  fixed by
recent experimental results from ELSA~\cite{Tran}. Furthermore, 
the photon illuminates the whole nucleus, so the reaction also 
occurs deep inside the nuclear interior, while  mesonic 
and baryonic projectiles
confine the reaction to the nuclear periphery~\cite{Bennhold1}.
Moreover, $K^+$-mesons only  weakly interact with the 
nuclear medium due to the strangeness conservation and thus the 
information about the production mechanism does not get distorted
significantly. This is essential, for example, for the study of 
in-medium self-energies of kaons that are presently being discussed
in the framework of heavy-ion collisions.

Historically~\cite{Rosental}, strangeness photoproduction 
from nuclei is considered as one of the optimal ways to study 
the formation of hypernuclei~\cite{Bennhold1,Bennhold2,Bennhold3,Abu}.
The hypernuclear production in $\gamma{A}$ reactions is part
of the program at the  Jefferson Laboratory~\cite{91016} as
well as the study of the $K^+$-meson
production mechanism in photo-nuclear reactions~\cite{91014}. 
It is the purpose of the present work to provide cross sections for
the photoproduction of kaons close to threshold to support  
later experimental and theoretical studies of
in-medium selfenergies of kaons. Since we work close to threshold
we are sensitive to details of the energy-momentum distribution in 
nuclei. We, therefore, use as in our earlier studies for proton induced
reactions~\cite{Sibirtsev3} state of the art nuclear 
dispersion relations~\cite{Sick}.
As a byproduct of our calculations we find that the spectral functions
of nucleons can also be obtained from the four-momenta of the produced 
hyperon and kaon if  their final state interactions are small.

\section{Elementary reaction}
To describe the production process in nuclear matter we need the
elementary $\gamma{p}{\to}K^+\Lambda$ reaction, which can be 
evaluated by the first order Feynman diagrams. Following the
pioneering studies by Gourdin and Dufour~\cite{Gourdin} or 
Thom~\cite{Thom} we include the Born terms in addition to 
low energy mesonic,  nucleonic and hyperonic resonances, with 
masses and widths fixed from the PDG~\cite{PDG} as 
listed in Table~\ref{tab1}. 
We take into account the contribution from $s$, $u$ and $t$ 
channels, but following Refs.~\cite{Gourdin,Thom} neglect 
form factors at the interaction vertices. 
For the current status of the kaon photoproduction operator 
we refer the reader to Refs.~\cite{Feuster1,Feuster2}. 
In the model of Refs.~\cite{Gourdin,Thom,Feuster1,Feuster2} the 
production amplitude is obtained as function of the energies and momenta 
separately of the incoming nucleons. This feature will later be 
used to explore the off-shell dependence of the cross section.

\begin{table}[h]
\caption{The particle properties from Ref.~\protect\cite{PDG} 
and the products of the photon and kaon coupling constants $G$ 
(in the notation of 
Ref.~\protect\cite{Williams}) used in our model.
The symbols $V$ and $T$ denote the vector and tensor coupling,
respectively.}
\vspace{2mm}
\label{tab1}
\renewcommand{\tabcolsep}{1.66pc} 
\renewcommand{\arraystretch}{1.3} 
\begin{tabular}{lllll}
\hline
Particle & $J^P $ & Mass  (MeV) & Width  (MeV) & G \\
\hline
$K^+$ & $0^-$ & 493.68 & & \\
$\Lambda$ & $\frac{1}{2}^+$ & 1115.68 & & 2.77$\pm$0.36 \\
$\Sigma$ &  $\frac{1}{2}^+$ & 1192.64 & & 9.23$\pm$0.44 \\
$K^\ast$  & $1^-$ & 891.66 & 50.8 & -0.39$\pm$0.12$^V$ \\
$K^\ast$   & & & & -0.001$\pm$0.18$^T$ \\
$K_1$ & $1^+$ & 1273 & 87 & 1.01$\pm$0.05$^V$ \\
$K_1$ & & & & 3.55$\pm$0.16$^T$ \\
$S_{11} $ & $\frac{1}{2}^-$ & 1650 & 150 & -0.417$\pm$0.010 \\
$P_{11}$ & $\frac{1}{2}^+$ & 1710 & 100 &  -0.962$\pm$0.065 \\
$S_{01}$ & $\frac{1}{2}^-$ & 1405 & 50 &  -0.488$\pm$0.15 \\
$S_{01}$ & $\frac{1}{2}^-$ &  1670 & 35 & 17.41$\pm$0.32 \\
$S_{01}$ & $\frac{1}{2}^-$ & 1800 & 300 &  0.07$\pm$0.33 \\
\hline
\end{tabular}\\[2pt]
\end{table}

Since we do not intend to extract the relevant coupling 
constants and  resonance properties, but to construct
simple parameterizations for the
$\gamma{p}{\to}K^+\Lambda$ differential cross sections,
we use the minimal set of free parameters as products of
photon and kaon couplings.

\begin{figure}
\vspace{-6mm}
\begin{minipage}[h]{80mm}
\psfig{file=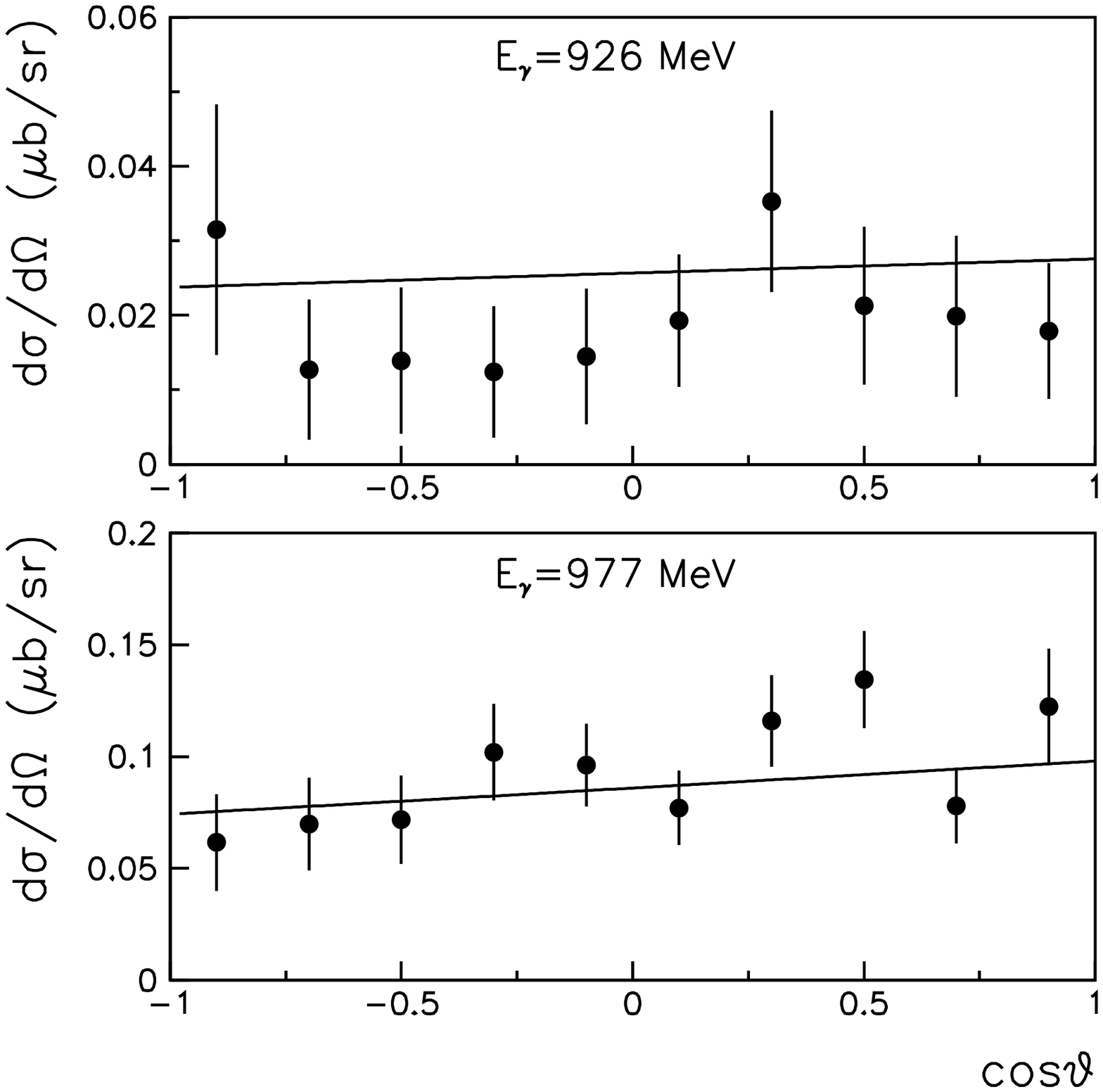,width=8.6cm,height=11.cm}
\end{minipage}
\phantom{aa}\hspace{-15mm}
\begin{minipage}[h]{80mm}
\psfig{file=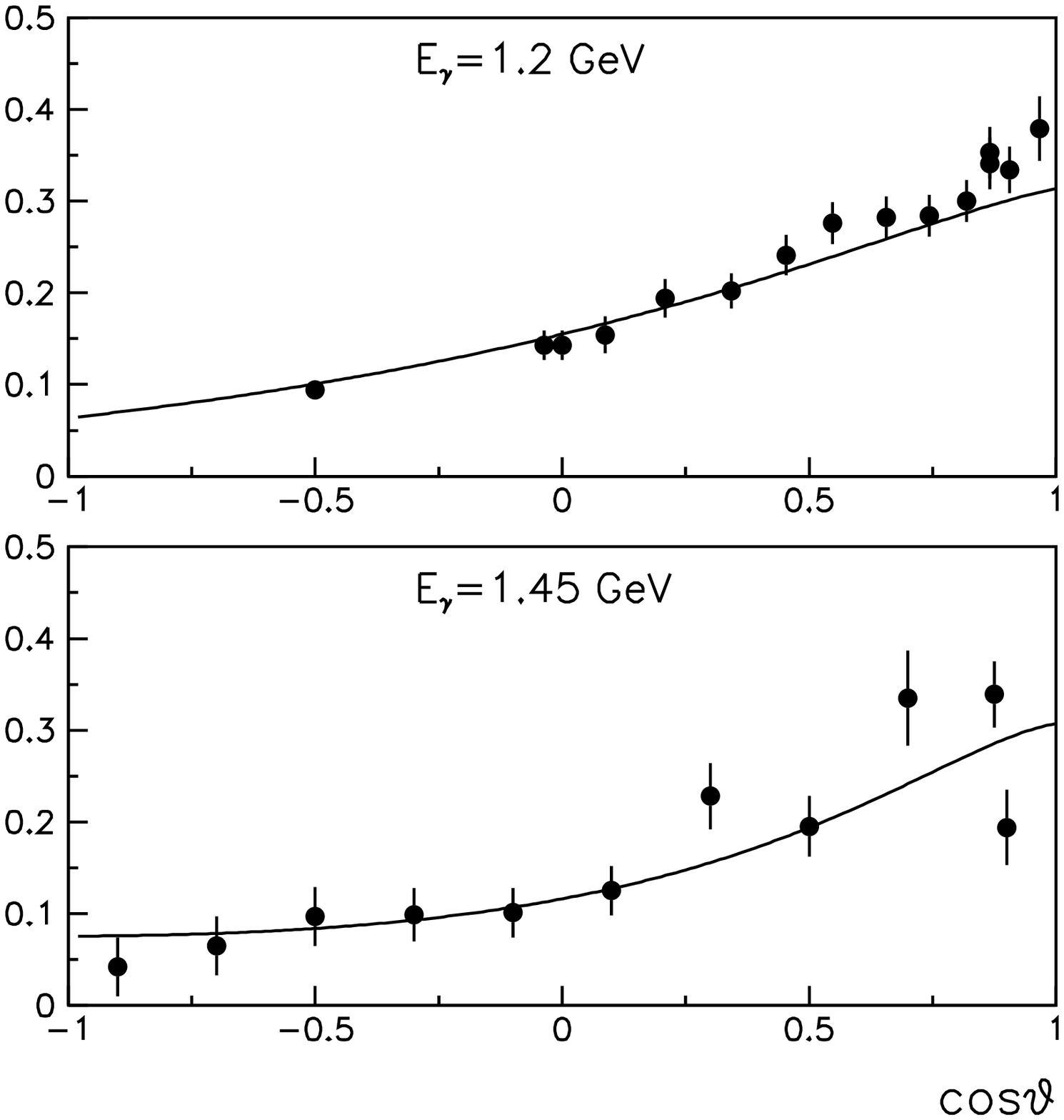,width=8.6cm,height=11.cm}
\end{minipage}
\phantom{aa}\vspace{-1.7cm}
\caption[]{The differential $\gamma{p}{\to}K^+\Lambda$
cross sections in the cm-system at four different photon 
energies $E_\gamma$. The experimental data are taken from 
Refs.~\protect\cite{Tran,LB} while the lines show
our fit.}
\label{gaka1}
\end{figure}

We fit the free parameters  to the available
data~\cite{Tran,LB} on the differential 
$\gamma{p}{\to}K^+\Lambda$ cross section 
at photon energies from 0.926~GeV to 2.1~GeV. Polarization data 
are not included in the fit. The fitting procedure  was 
performed by Minuit~\cite{James} and provides a reduced
$\chi^2{/}N$=3.88. The parameters $G$ are listed in Table~\ref{tab1}
where the notation for the coupling constants is that of 
Ref.~\cite{Williams}. Fig.~\ref{gaka1} shows the experimental 
data~\cite{Tran,LB} on the differential $\gamma{p}{\to}K^+\Lambda$ 
cross section in the cm-system for $E_\gamma$ from 0.926
to 1.45 GeV whereas  the solid lines indicate our parameterization.

\section{$K^+$ distortion in nuclear matter}
\begin{figure}[h]
\vspace{-8mm}
\phantom{aa}\hspace{-7mm}
\begin{minipage}[t]{80mm}
\psfig{file=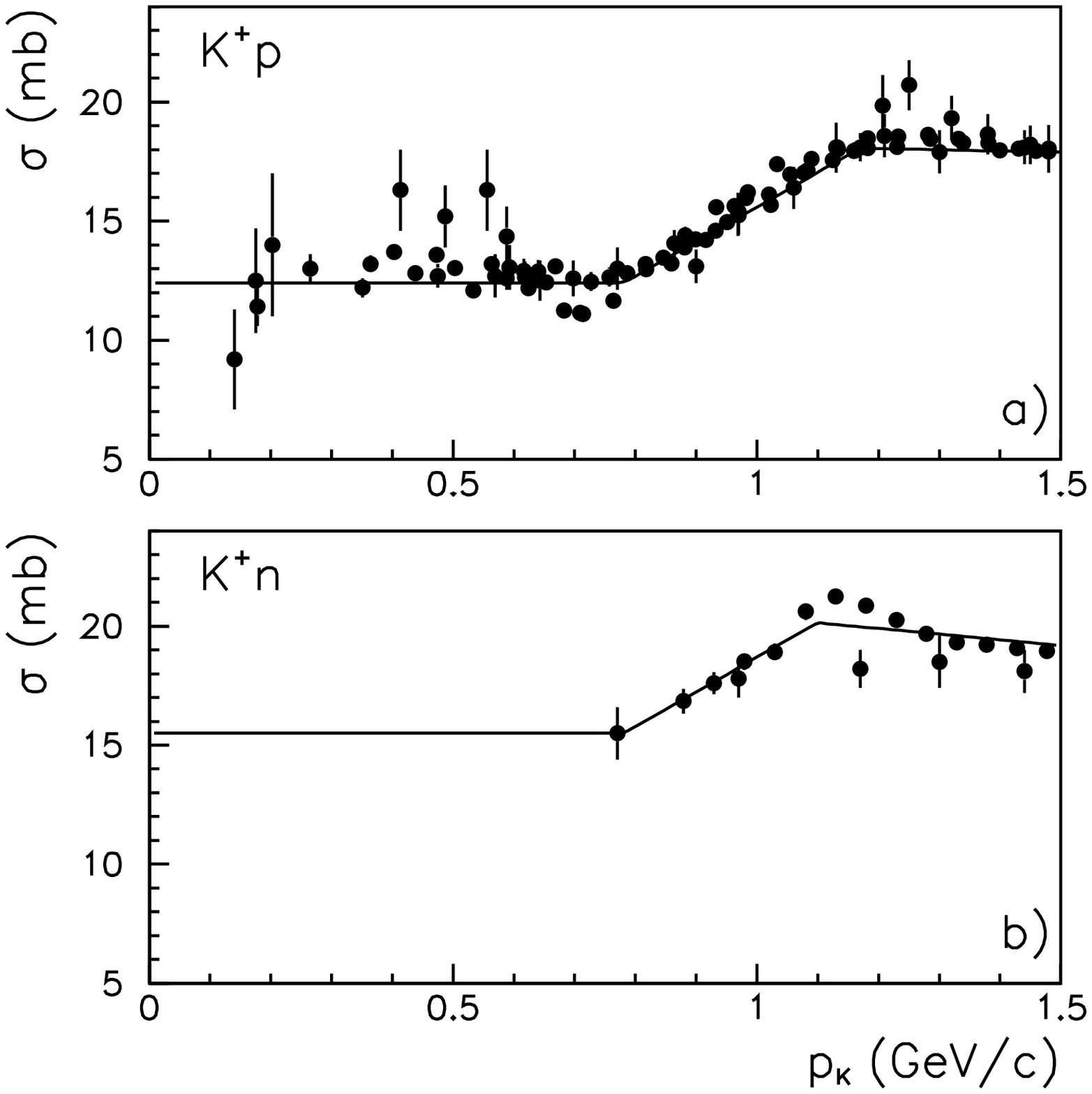,width=8.6cm,height=10cm}
\end{minipage}
\phantom{aa}\hspace{-15mm}
\begin{minipage}[t]{80mm}
\psfig{file=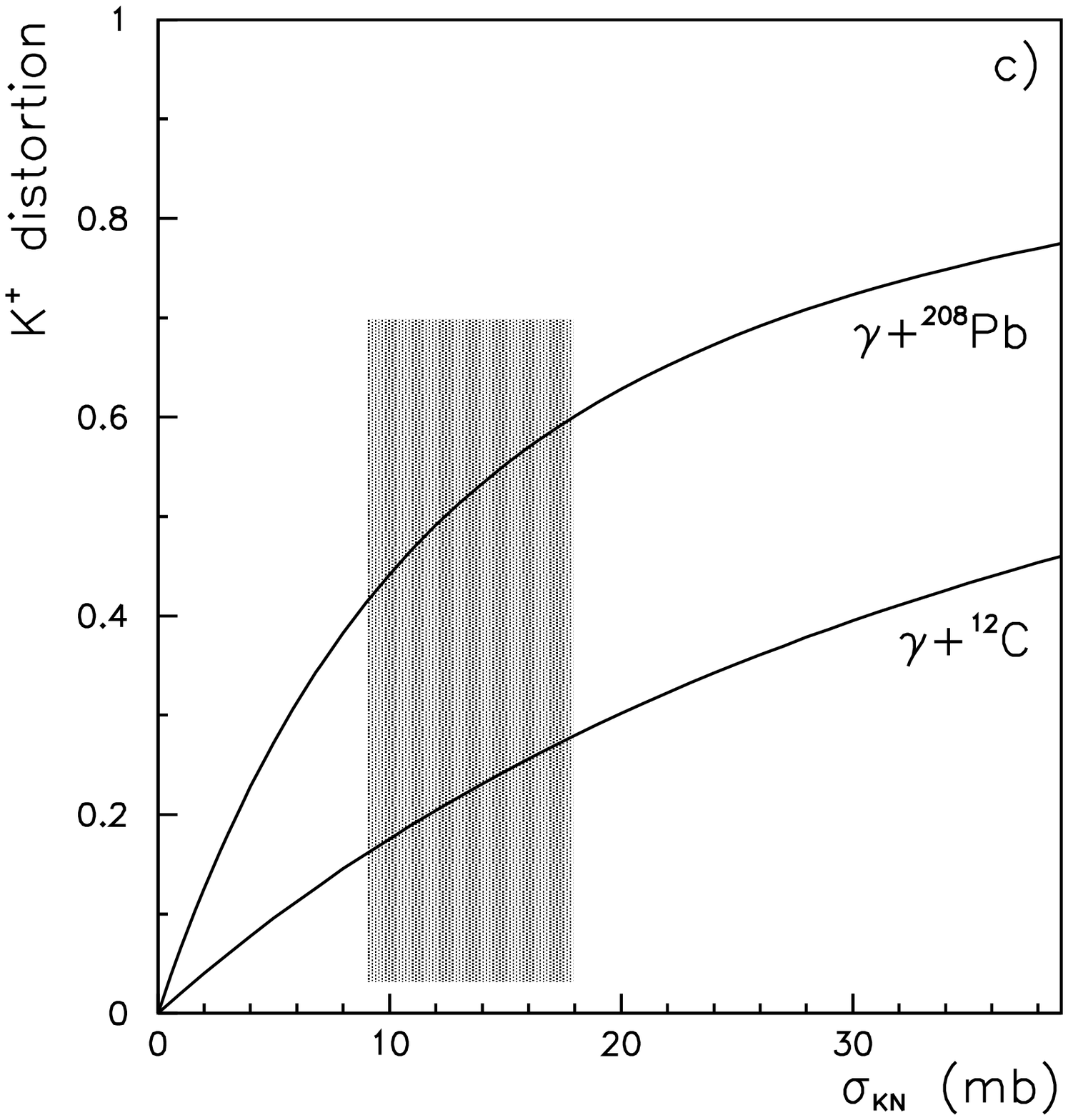,width=8.6cm,height=10cm}
\end{minipage}
\phantom{aa}\vspace{-1.7cm}
\caption[]{Total $K^+p$ (a) and $K^+n$ (b) cross sections as 
a function of the $K^+$-meson momentum from 
Ref.~\protect\cite{PDG}. The solid lines in (a,b) show the
parameterizations from Ref.~\protect\cite{Sibirtsev1}. (c) The
$K^+$ distortion factor $\kappa$ calculated in the Glauber model
for $\gamma{C}$ and $\gamma{Pb}$ as a 
function of the $K^+$-nucleon cross section. The dashed area indicates
$\sigma_{KN}$ in vacuum for kaon momenta below 1~GeV.}
\label{gaka45}
\end{figure}

Because of the rather weak $K^+$-nucleon interaction the $K^+$-meson
is considered as a promising probe for studying the production process
in  nuclear matter. Experimental data~\cite{PDG} on $K^+p$ and 
$K^+n$ cross sections are shown in  Fig.~\ref{gaka45}a-b)
together with the parameterizations from Ref.~\cite{Sibirtsev1}.
Note, that  data are not available for the $K^+n$ cross 
section at momenta below 800~MeV/c and the parameterization was
constructed in line with the low energy limit from the J\"ulich
model~\cite{Hoffman}. Within the experimental uncertainties
the $K^+N$ cross sections range from 9  to 18~mb at kaon
momenta below 1~GeV/c. Therefore, at normal  nuclear density
the kaon mean-free-path is  within the range of 
$\simeq$3.5$\div$7~fm,  comparable to a typical nuclear 
size. Because of the large mean-free-path it is expected that
$K^+$-mesons should be only slightly distorted especially
for light targets.  

A measure for the final state interaction is provided by
the distortion factor $\kappa$, calculated within the Glauber
model as
\begin{equation}
\kappa = 1 - \frac{A_{eff}(\sigma_{KN})}{A_{eff}(\sigma_{KN}=0)}.
\label{distort}
\end{equation}
Here $A_{eff}$ is the effective number of nucleons participating in
the production and propagation of the kaon, which in case of 
photoproduction is given as~\cite{Kolbig,Benhar,Vercellin,Effenberger}
\begin{equation}
A_{eff}(\sigma_{KN}) = \int\limits_0^{+\infty} d{\bf b}  
\int\limits_{-\infty}^{+\infty} dz \ {\rho}({\bf b},z) \
\exp [ -{\sigma}_{KN}
\int\limits_z^{\infty} d\xi \ {\rho}({\bf b}, \xi) ],
\end{equation}
with $\rho({\bf r})$ denoting the single-particle density,
that was taken of  Fermi type with parameters from 
Ref.~\cite{Jager}. The density function is normalized such that
$A_{eff}(\sigma_{KN}{=}0){=}A$.

Fig.~\ref{gaka45}c) shows the $K^+$ distortion factor $\kappa$ as
a function of the $K^+N$ cross section  for $\gamma{C}$ and
$\gamma{Pb}$ interactions. The dashed area in  Fig.~\ref{gaka45}c)
indicates the cross section $\sigma_{KN}$ in vacuum for 
kaon momenta below 1~GeV/c. Even for the carbon target there 
is a roughly  20\% probability for the $K^+$-meson to interact 
in the nuclear interior. For the $Pb$ nucleus the $K^+$ distortion 
is quite substantial. 
Because of  strangeness conservation there are no  inelastic 
channels apart from the $K^+n{\to}K^0p$ charge exchange reaction
leading to $K^+$ absorption. The low energy $K^+N$
interaction is entirely due to  elastic scattering such that  
the $K^+$ distortion factor provides the probability for 
$K^+N$ elastic scattering in nuclear matter.

Obviously, the $K^+N{\to}K^+N$ scattering distorts the information
about the kaon production mechanism in nuclear matter and thus 
heavy nuclei are not  suited for a relevant study. The
elastic scattering populates the low energy part of the kaon
spectrum, which can be also understood in terms of a
thermalization process~\cite{Sibirtsev2}.
Thus, to reconstruct the production mechanism,  light
nuclear targets are an optimal choice and in the 
following we will use $^{12}C$. Moreover, since the total
$K^+$-meson flux is almost conserved (when neglecting the small
absorption due to the $K^+n{\to}K^0p$ charge exchange)
the $A$-dependence of the total $K^+$ production cross section
can deviate from a linear scaling with $A$ only due to 
the production process, which is not necessarily  equivalent for
light and heavy nuclei.

\section{Dispersion relation in nuclei}
\begin{figure}[h]
\vspace{-8mm}
\phantom{aa}\hspace{-7mm}
\begin{minipage}[t]{80mm}
\psfig{file=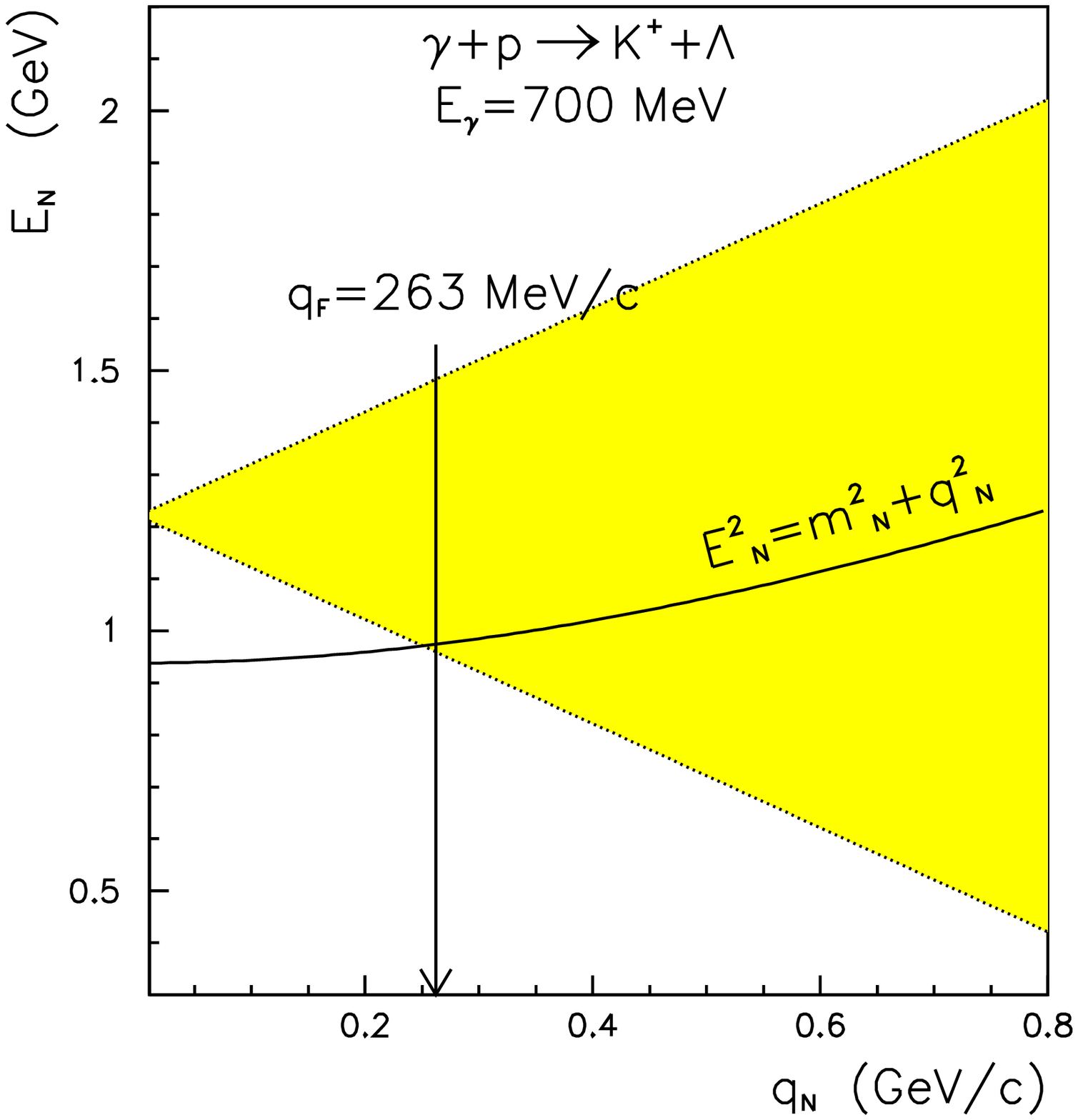,width=8.6cm,height=10cm}
\end{minipage}
\phantom{aa}\hspace{-15mm}
\begin{minipage}[t]{80mm}
\psfig{file=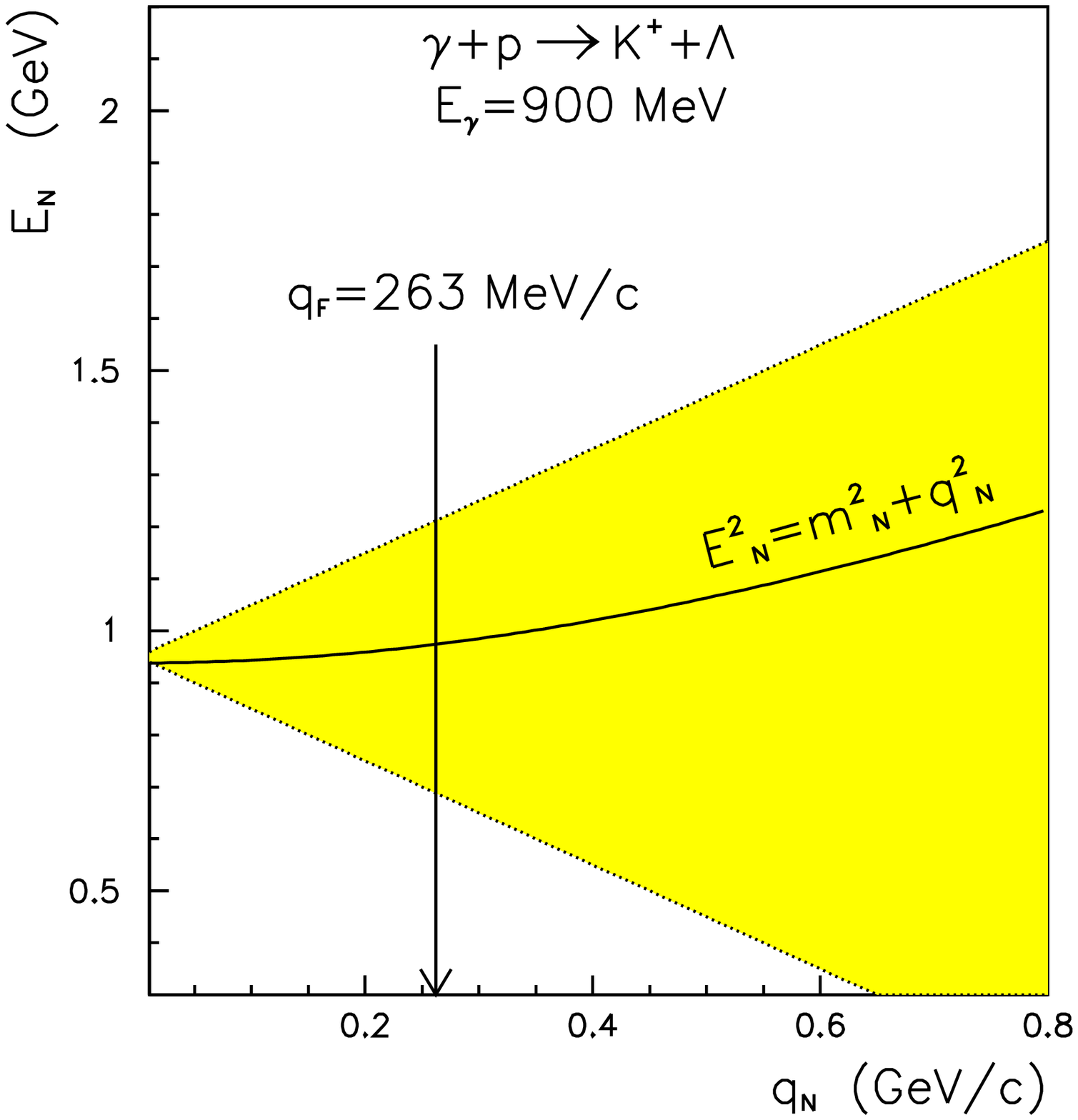,width=8.6cm,height=10cm}
\end{minipage}
\phantom{aa}\vspace{-1.7cm}
\caption[]{The energetically possible range (dashed area) 
of  nucleon momenta $q_N$ and
energies $E_N$  for the $\gamma{p}{\to}K^+\Lambda$
reaction at photon energies of 700~MeV and 900~MeV. The solid line 
indicates the dispersion relation in free space while the arrow
shows the Fermi momentum for nuclear matter.}
\label{gaka20}
\end{figure}

The $\gamma{p}{\to}K^+\Lambda$ reaction threshold is 
given by energy conservation as
\begin{equation}
\left( m_K +m_\Lambda \right)^2 = \left(E_\gamma +E_N\right)^2 -
\left( {\bf k}_\gamma + {\bf q}_N \right)^2,
\label{conserv}
\end{equation}
where in nuclear matter $q_N{\neq}0$ and
$E_N{\neq}m_N$. In principle, energy conservation provides
the spectrum of nucleon momenta $q_N$ and energies $E_N$
available for $K^+\Lambda$ production, which depends on
the photon energy. The dashed areas in Fig.~\ref{gaka20}
indicate the range of  $q_N$ and $E_N$  that can energetically
contribute to strangeness production at 
$E_\gamma$=700 and $E_\gamma$=900~MeV. The free 
dispersion relation $E_N{=}\sqrt{q_N^2{+}m_N^2}$  is shown 
in Fig.~\ref{gaka20} by the solid line.

Using the free dispersion relation Fig.~\ref{gaka20}
illustrates that for $K^+\Lambda$ production at a
photon energy of 700~MeV one needs nucleon momenta 
above $\simeq$260~MeV/c, which are only barely available
in the Fermi sea. Note that the Fermi momentum for 
nuclear matter is $q_F{\simeq}$263~MeV/c, which is shown 
in Fig.~\ref{gaka20} by an arrow.  On the other hand,  
at $E_\gamma$=900~MeV the  $K^+$-meson can be  produced almost
with $q_N{\simeq}0$, which  reflects the fact that the 
$\gamma{p}{\to}K^+\Lambda$ reaction threshold lies at a
photon energy of $\simeq$911~MeV in free space.
The kinematical considerations shown in  Fig.~\ref{gaka20}
motivate our present study of  particle production at energies
below the reaction threshold in vacuum. Indeed, it is clear 
that at photon energies below 900~MeV the $K^+\Lambda$ production
becomes sensitive to nucleon momenta $q_N{>}q_F$ that
may be attributed to the high momentum component of the nuclear wave 
function.

The situation is even  exciting since in nuclear matter
the free dispersion relation does not hold. There are different
models~\cite{Atti,Benhar1,Benhar2,Sick} providing 
a relation between the nucleon momentum
$q_N$ and energy $E_N$ in nuclei, but all of them correspond
to an  $E_N(q_N)$ spectrum below the free dispersion relation,
i.e. below the solid line shown in Fig.~\ref{gaka20}.

Furthermore, the dispersion relations used for  particle 
production in hadron-nucleus reactions cannot be uniquely fixed 
by theory; we thus suggest to study this important aspect of
production dynamics  by
$\gamma{A}{\to}K^+\Lambda{X}$ reactions. To illustrate the 
sensitivity of observables to  different dispersion
relations we perform  model estimates using 
two well-known approaches, i.e.  the Thomas-Fermi approximation 
and  the spectral function. 

\section{Thomas-Fermi approximation}
In the noninteracting Fermi-gas model the occupation probability 
of states with momentum $q_N$ is given by
\begin{equation}
\Phi (q_N) = \frac{3} {4 \pi q_F^{3/2}} 
\Theta (q_F- |{\bf q}_N| ),
\end{equation}
where $q_F$ is the Fermi momentum;  for nuclear matter it
is related to the average nuclear density 
$\rho_0$=0.16~fm$^{-3}$  by
\begin{equation}
q_F = \left( \frac{3 \pi^2 \rho_0}{2} \right)^{1/3}.
\end{equation}

\begin{figure}[h]
\vspace{-6mm}
\begin{minipage}[t]{80mm}
\psfig{file=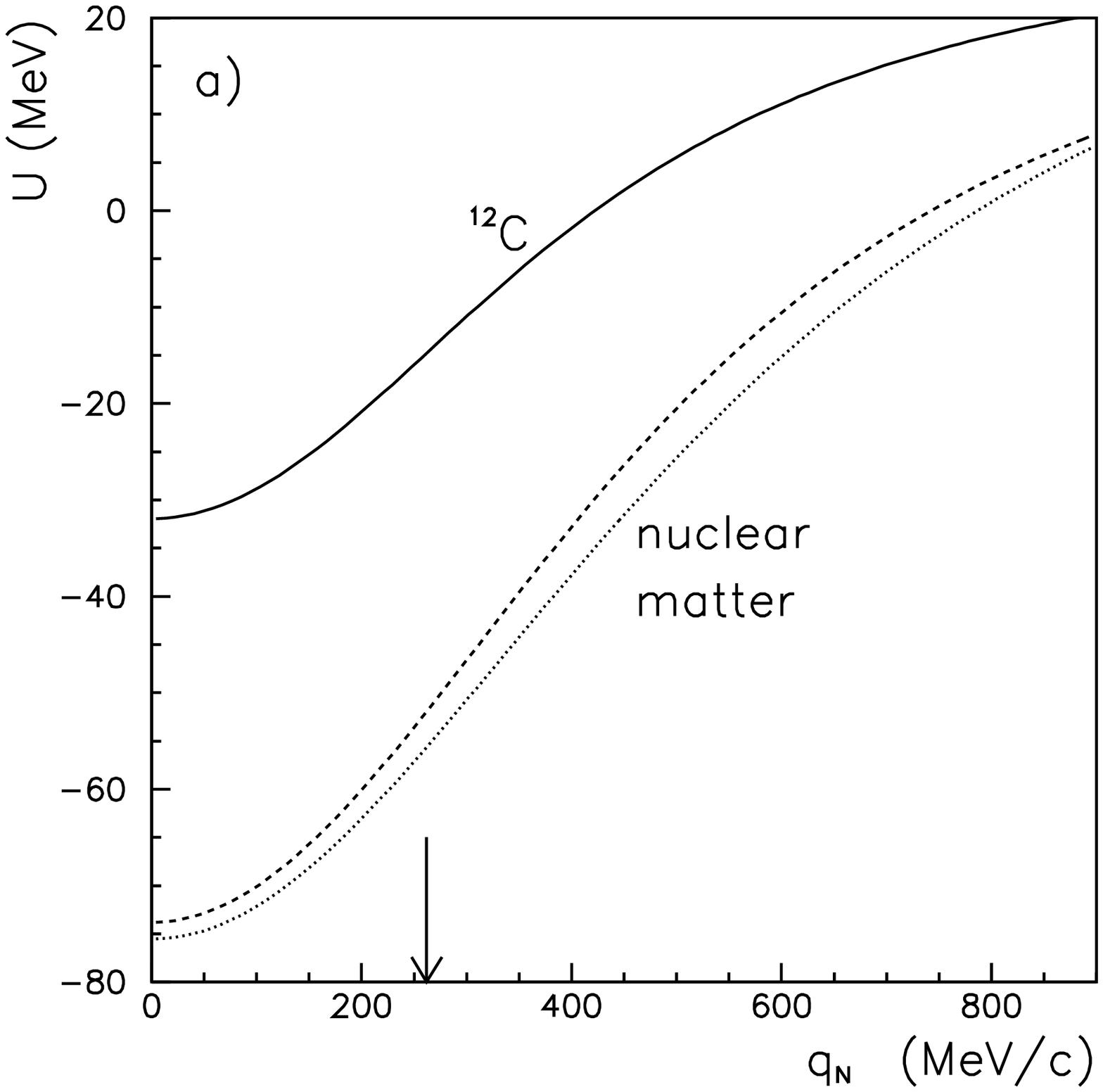,width=8.cm,height=10cm}
\end{minipage}
\begin{minipage}[t]{80mm}
\psfig{file=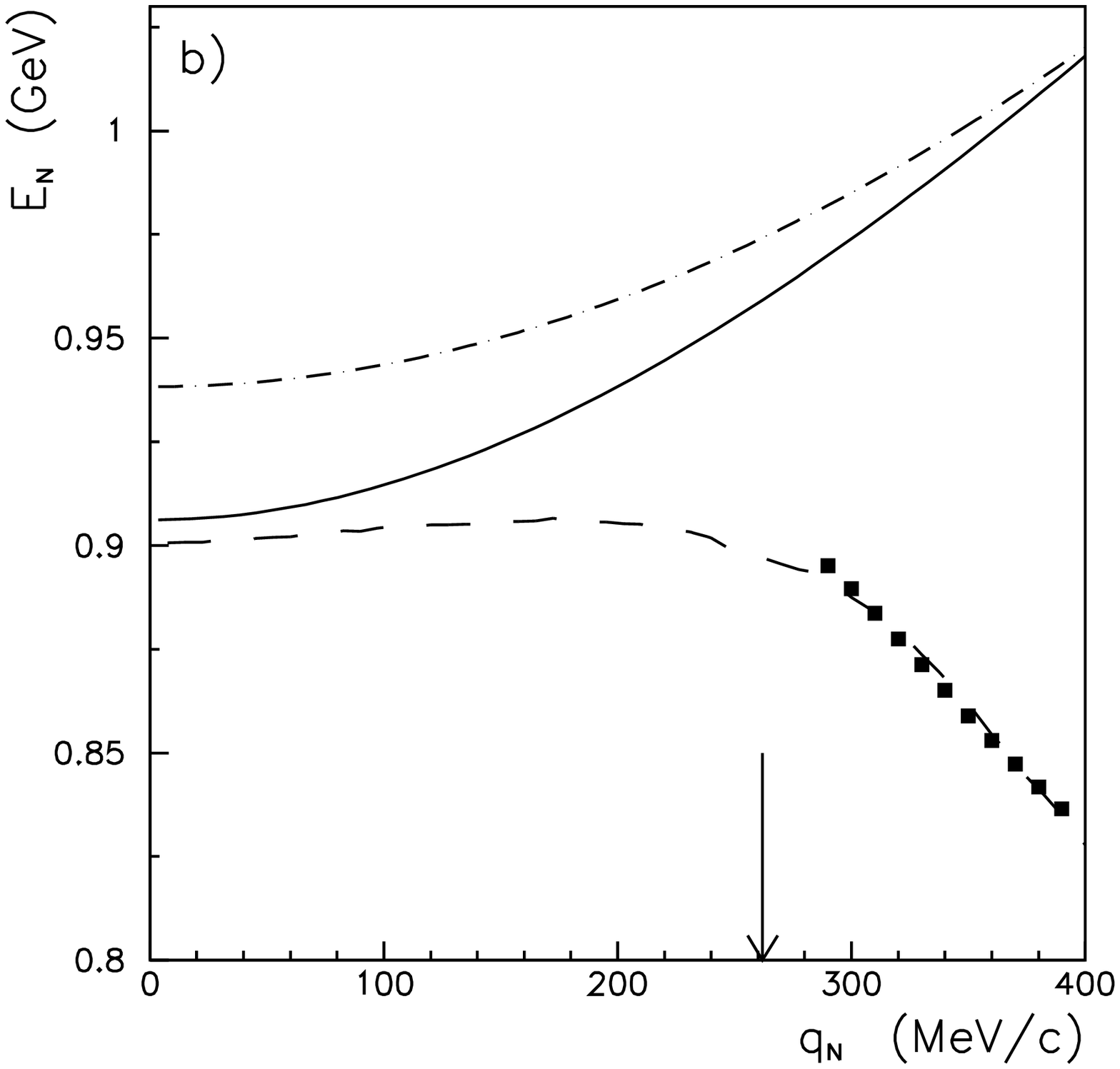,width=8.cm,height=10cm}
\end{minipage}
\phantom{aa}\vspace{-15mm}
\caption{a) The single-nucleon potential at 
$\rho$=0.16~fm$^{-3}$
as a function of the nucleon momentum $q_N$. The results for nuclear
matter are shown by the dashed~\protect\cite{Welke}  and
the dotted line~\protect\cite{Wiringa}. The solid line shows the 
potential for $^{12}C$ calculated by Eq.~(\protect\ref{pwelke})
with $q_F$=221~MeV/c. The
arrow indicates the Fermi momentum for nuclear matter.
b) The dispersion relation $E_N(q_N)$ for $^{12}C$. The 
solid  line shows the relation~(\ref{disp1}) with the 
potential~(\protect\ref{pwelke}), the dash-dotted  line is the free 
dispersion relation $E_N{=}\sqrt{q_N^2{+}m_N^2}$ while
the long-dashed line shows the calculation  with the spectral function.
The squares display the result for the correlated part
of the spectral function calculated by the model proposed
in Ref.~\protect\cite{Atti}.}
\label{gaka8}
\vspace{-2mm}
\end{figure}

The Fermi momenta for finite nuclei are given in 
Ref.~\cite{Frullani}; for $^{12}C$ we have $q_F$=221~MeV/c. 
Within the local Tomas-Fermi approximation the total energy of 
the struck nucleon then is
\begin{equation}
E_N = \sqrt{q_N^2+m_N^2} + U(q_N,\rho),
\label{disp1}
\end{equation}
where $m_N$ is the free nucleon mass and the nuclear potential 
$U$ depends on the nucleon momentum $q_N$ and density $\rho$.
For the following calculations we adopt the
potential given by Gale, Bertsch and Das Gupta~\cite{Gale} and
Welke et al.~\cite{Welke}
for cold nuclear matter as
\begin{eqnarray}
U(q_N,\rho) = A_1\frac{\rho}{\rho_0} + 
B_1\left(\frac{\rho}{\rho_0}\right)^{1.24} +
\frac{2 \pi C_1 \Lambda_1}{\rho_0}
\left[
\frac {q_F^2 + \Lambda_1^2 -q_N^2} {2 q_N \Lambda_1} \times
\ln \frac {(q_N+q_F)^2 +\Lambda_1^2} {(q_N-q_F)^2 +\Lambda_1^2}
\right. \nonumber \\ \left. + \frac {2q_F } {\Lambda_1 }-
2 \left( arctg \frac {q_N+q_F} {\Lambda_1} 
- arctg \frac {q_N-q_F} {\Lambda_1} \right) \right],
\label{pwelke}
\end{eqnarray}
where $A_1$=--110.44~MeV, $B_1$=140.9~MeV, $C_1$=-64.95~MeV 
and $\Lambda_1$=1.58$q_F$.
The dashed line in Fig.~\ref{gaka8}a) shows the 
potential~(\ref{pwelke}) for nuclear matter with $q_F$=262~MeV/c 
at $\rho{=}\rho_0$ as a function of the nucleon momentum. 
Note, that the mean field without momentum dependence provides 
$U{\simeq}$--53~MeV~\cite{Bertch} for nuclear matter. 
The analysis of  experimental scattering data using 
the Dirac equation~\cite{Weber,Cooper,Hama}
indicates that the potential becomes repulsive at nucleon
energies $E_{kin}$ above $\simeq$300~MeV. 

The potential~(\ref{pwelke})  for  $^{12}C$ ( using
$q_F$=221~MeV/c ) is shown by the solid line in 
Fig.~\ref{gaka8}a). In its ground state  $^{12}C$  has 
$s_{1/2}$ and $p_{3/2}$ single-particle levels~\cite{Greiner}
with single-particle energies of $\simeq$--34 and --16~MeV, respectively. 
Indeed,  Fig.~\ref{gaka8}a) indicates that the potential~(\ref{pwelke})  
approaches the lowest single-particle energy at low 
nucleon momenta.

Another optical potential was proposed by Wiringa~\cite{Wiringa} as
\begin{equation}
U(q_N,\rho) = A_2\frac{\rho}{\rho_0} + 
B_2\left(\frac{\rho}{\rho_0}\right)^2+
C_2\frac{\rho}{\rho_0}\left(
1+[q_N/\Lambda_2]^2 \right)^{-1},
\label{pwiringa}
\end{equation}
where $A_2$=15.52~MeV, $B_2$=24.93~MeV, $C_2$=-116~MeV and
$\Lambda_2{=}3.29{-}0.373(\rho/\rho_0)$ 
fm$^{-1}$~\cite{Pandharipande}. The dashed line in
Fig.~\ref{gaka8}a) shows the potential~(\ref{pwiringa})
for  nuclear matter at density $\rho{=}0.16$~fm$^{-3}$, 
which at nucleon momenta
below $q_F$ (for density $\rho_0$ ) provides  almost 
identical results as (\ref{pwelke}).

The dispersion relation for $^{12}C$ is given
by Eq.~(\ref{disp1}) and  shown in Fig.~\ref{gaka8}b)
by the solid line while the  dash-dotted line indicates the 
free dispersion relation $E_N{=}\sqrt{q_N^2{+}m_N^2}$.
For low nucleon momenta $q_N{<}$100~MeV/c the Thomas-Fermi model 
with the Carbon potential is in rough agreement with the result 
from the spectral function approach of Ref.~\cite{Sick,Sick1} shown 
in Fig.~\ref{gaka8}b) by  the long-dashed line. 

\section{The spectral function}
\begin{figure}[h]
\phantom{aa}\vspace{-8mm}
\begin{minipage}[t]{80mm}
\psfig{file=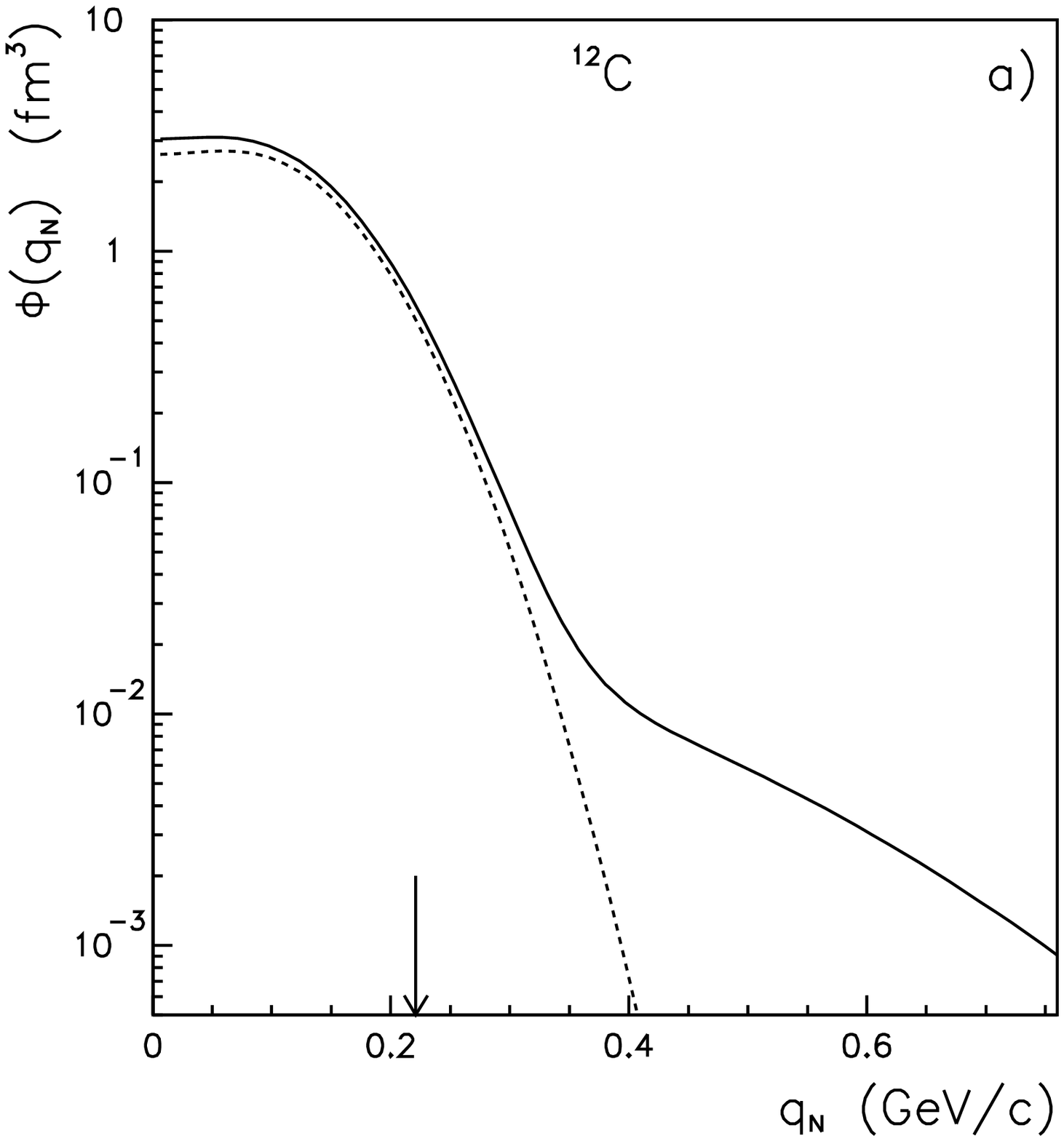,width=8.cm,height=10cm}
\end{minipage}
\begin{minipage}[t]{80mm}
\psfig{file=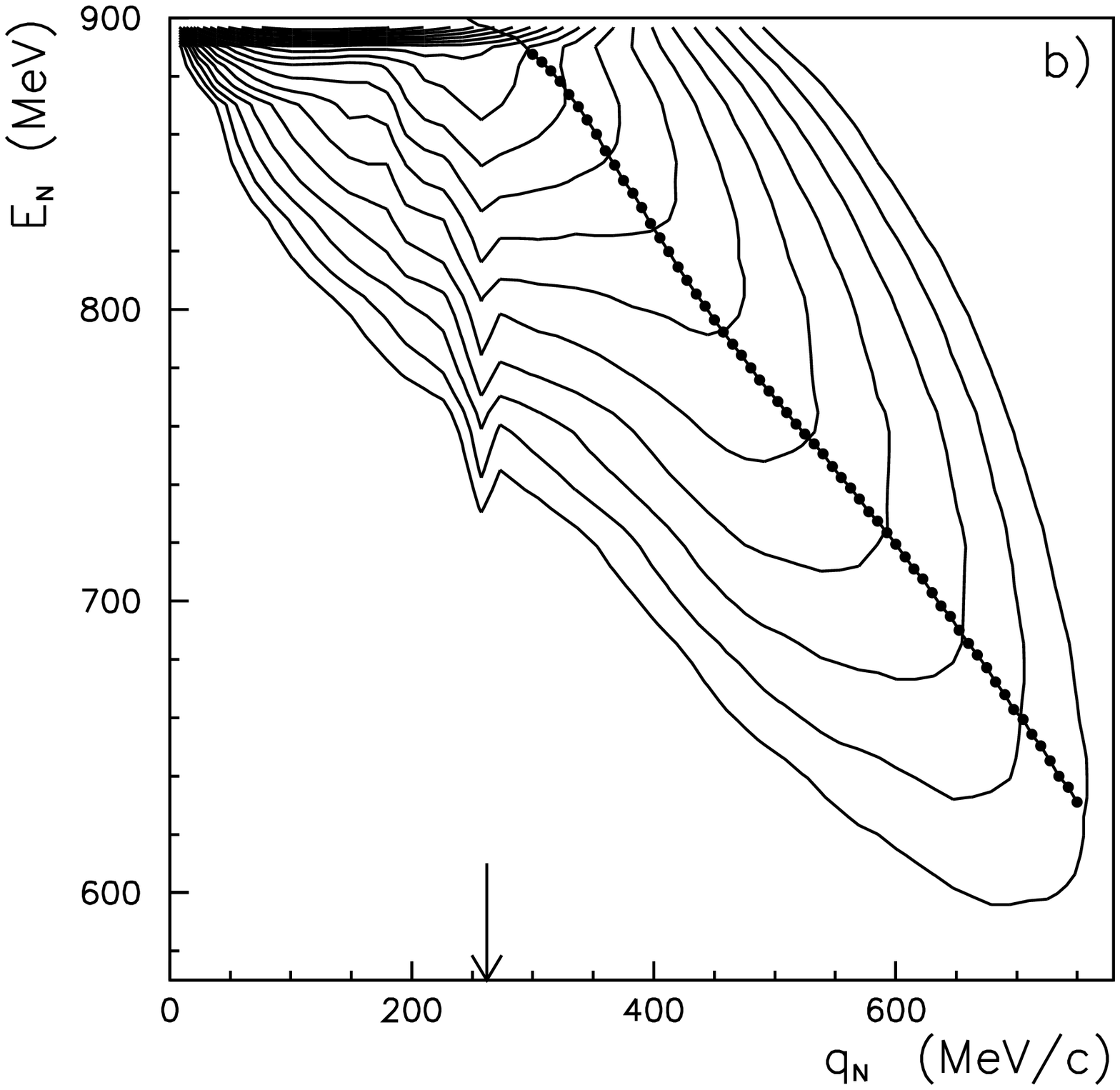,width=8.cm,height=10cm}
\end{minipage}
\phantom{aa}\vspace{-15mm}
\caption{a) The nucleon momentum distribution $\Phi{(}q)$
for $^{12}C$ taken from Ref.~\protect\cite{Atti}. 
The dashed line shows the uncorrelated part, while the
solid line is the sum of the uncorrelated and correlated parts.
The arrow indicates the Fermi momentum for $^{12}C$ from
Ref.~\protect\cite{Frullani}. b) The dispersion relation
given by the spectral function from Refs.~\cite{Sick,Sick1}.
The circles indicate the average nucleon energy calculated by 
Eq.~(\protect\ref{aver}).}
\label{gaka3}
\vspace{-2mm}
\end{figure}

The short-range $NN$ interactions induce correlations in the
nuclear wave function, which deplete states below the Fermi 
sea and partially occupy the states above the Fermi 
level~\cite{Benhar1,Fantoni}. Thus the nuclear momentum
distribution has a tail for  momenta $q_N{>}q_F$. 
The dashed line in Fig.~\ref{gaka3}a) shows the 
momentum distribution $\Phi{(}q)$
from Ref.~\cite{Atti} due to the uncorrelated part 
of the spectral function, while the solid line indicates the 
total momentum distribution including the two-body correlations. 
Although the contribution from the
high momentum component to the total function $\Phi{(}q)$
is small, its role can be important, since high nucleon 
momenta $q_N$ might provide the large energies required 
for particle production below the reaction threshold in free space.    

A nucleon in nuclear matter does not have a  sharp 
energy $E_N$  that is  uniquely defined by its momentum
$q_N$. The relation between the momentum and energy instead is given by 
the spectral function $S(q_N,E_N)$, which provides the joint
probability of finding a  nucleon with momentum
$q_N$ and energy $E_N$. The spectral function is generally
given~\cite{Benhar2,Sick,Mahaux,Baldo,Atti2,Cordoba} in terms 
of the removal energy $E_R$, which is  defined by 
energy conservation in the dissociation of nucleus $A$ into
nucleus $B$ and a single nucleon as~\cite{Frullani}
\begin{equation}
M_A = M_B+E^\ast_B   +
\left( m_N - E_R \right),
\label{dissip}
\end{equation}
where $M_A$ and $M_B$ are the masses of the nuclei $A$ and $B$ 
in their ground state and $E^\ast_B$ is the excitation energy 
of the residual system $B$. Due to momentum conservation
both, the residual system and nucleon, have the same modulus of
the three-momentum $q_N$. In Eq.~(\ref{dissip}) the term
in the brackets stands for the total nucleon energy
and thus we have
\begin{equation}
E_N = m_N -E_R.
\end{equation} 

In the following calculations we use the spectral 
function for $^{12}C$ as given by Sick et al.~\cite{Sick,Sick1}, 
which is shown 
in Fig.~\ref{gaka3}b). Indeed, within the dispersion relation
given by Fig.~\ref{gaka3}b) a  certain nucleon momentum $q_N$
is now related to the spectrum of the nucleon energy $E_N$.
The contribution from $s_{1/2}$ and $p_{3/2}$ single-particle
levels  is indicated by the weak ridge at $q{<}$300~MeV/c, while 
the spectrum beyond the shell structure stems
from the short-range and tensor correlations. It is important
to note that on the average  large momenta $q_N{>}$300~MeV/c are 
associated with  small values of the nucleon energy $E_N$.
The dependence of the average nucleon energy $<E_N>$ 
on the momentum $q_N$ is given as
\begin{equation}
<E_N> = \frac {1} {\Phi (q_N)} 
\int\limits_{E_{th}}^\infty E_N \ S(q_N,E_N) \ dE_N, 
\label{aver}
\end{equation}
while the momentum distribution is related to  the spectral 
function as
\begin{equation}
\Phi (q_N) = \int\limits_{E_{th}}^\infty S(q_N,E_N) \ dE_N, 
\end{equation}
where $E_{th}$ is the single nucleon removal threshold.
The average nucleon energy is shown in Fig.~\ref{gaka3}b)
by the dots and indicates that  high nucleon momenta 
correspond to high removal energies. The dependence of ${<}E_N{>}$ 
on the momentum $q_N$ can be compared to the dispersion
relation given by the Thomas-Fermi model as  shown
in Fig.~\ref{gaka8}b). The result from the spectral 
function (solid line) is  close to the Thomas-Fermi
approach at $q_N{<}$100~MeV/c, but the two models differ 
substantially  at larger momenta. Notice that even
below the Fermi momentum $q_F{=}$221~MeV/c the difference 
is quite strong.

The squares in Fig.~\ref{gaka8}b) indicate the average nucleon
energy  $<E_N>$ resulting from the correlated part of the
$^{12}C$ spectral function as calculated in the model proposed by
Ciofi degli Atti and Simula~\cite{Atti}. As  can be seen,
the agreement between the different models for the spectral
function is very reasonable.

\section{Results for $K^+$ production in $\gamma{A}$ reactions}
The Lorentz-invariant differential cross section for 
$K^+$-meson production in $\gamma{A}$
reaction is given  as~\cite{Sibirtsev3}
\begin{equation}
E_K \frac{d^3\sigma_{\gamma A\to K \Lambda X} }{d^3p_K} =
A_{eff}(\sigma_{KN}) 
\int  d^3 q_N \ dE_N \ S(q_N, E_N) \ \frac{Z}{A} \ E_K^{\prime}
\frac{d^3 \sigma_{\gamma p\to K^+ \Lambda}(\sqrt{s},{\cal P}_N)}
{d^3p_K^{\prime}},
\label{direct}
\end{equation}
where $\sqrt{s}$ is the invariant energy of the incident photon 
and the struck nucleon defined as
\begin{equation}
s = (E_\gamma +E_N)^2 -({\bf k}_\gamma + {\bf q}_N)^2,
\end{equation}
${\cal P}_N{=}(E_N,{\bf q}_N)$ denotes the four-momentum of the
struck nucleon
and the limits of integration in Eq.~(\ref{direct}) are fixed 
by energy conservation as
\begin{equation}
m_K +m_\Lambda \le \sqrt{s} \le \sqrt{2 E_\gamma M_A +
M_A^2} -M_{A-1}
\label{limits}
\end{equation} 
with $M_A$ and $M_{A-1}$ denoting the mass of the initial
and final $A{-}1$ nuclei, respectively. Note, that the
upper limit in~(\ref{limits}) corresponds to the
ground state of the residual $A{-}1$ nucleus and provides the maximal
available energy $\sqrt{s}$ of the interaction vertex.

The  primed indices in Eq.~(\ref{direct}) 
denote the $K^+$-meson momentum and energy in 
the cms of the photon and the individual target nucleon, while
$E_K' d^3\sigma/d^3p_K'$ for 
${\cal P}_N{=}(\sqrt{q_N^2{+}m_N^2},{\bf q}_N)$
is the Lorentz invariant elementary  
cross section for the $\gamma{p}\to{K^+}\Lambda$ reaction in free space
as described in Sect.~2. The factor $Z/A$ accounts
for the fact that the reaction takes place only on the target protons.

The factor $A_{eff}$ stands for the effective primary collision  
number as well as for the in-medium distortion of the 
$K^+$-meson when propagating through the nuclear target as described
in Sect.~3. Note, that $A_{eff}{=}A$ when neglecting 
the kaon distortion. 

\begin{figure}[h]
\phantom{aa}\vspace{-8mm}
\begin{minipage}[t]{80mm}
\psfig{file=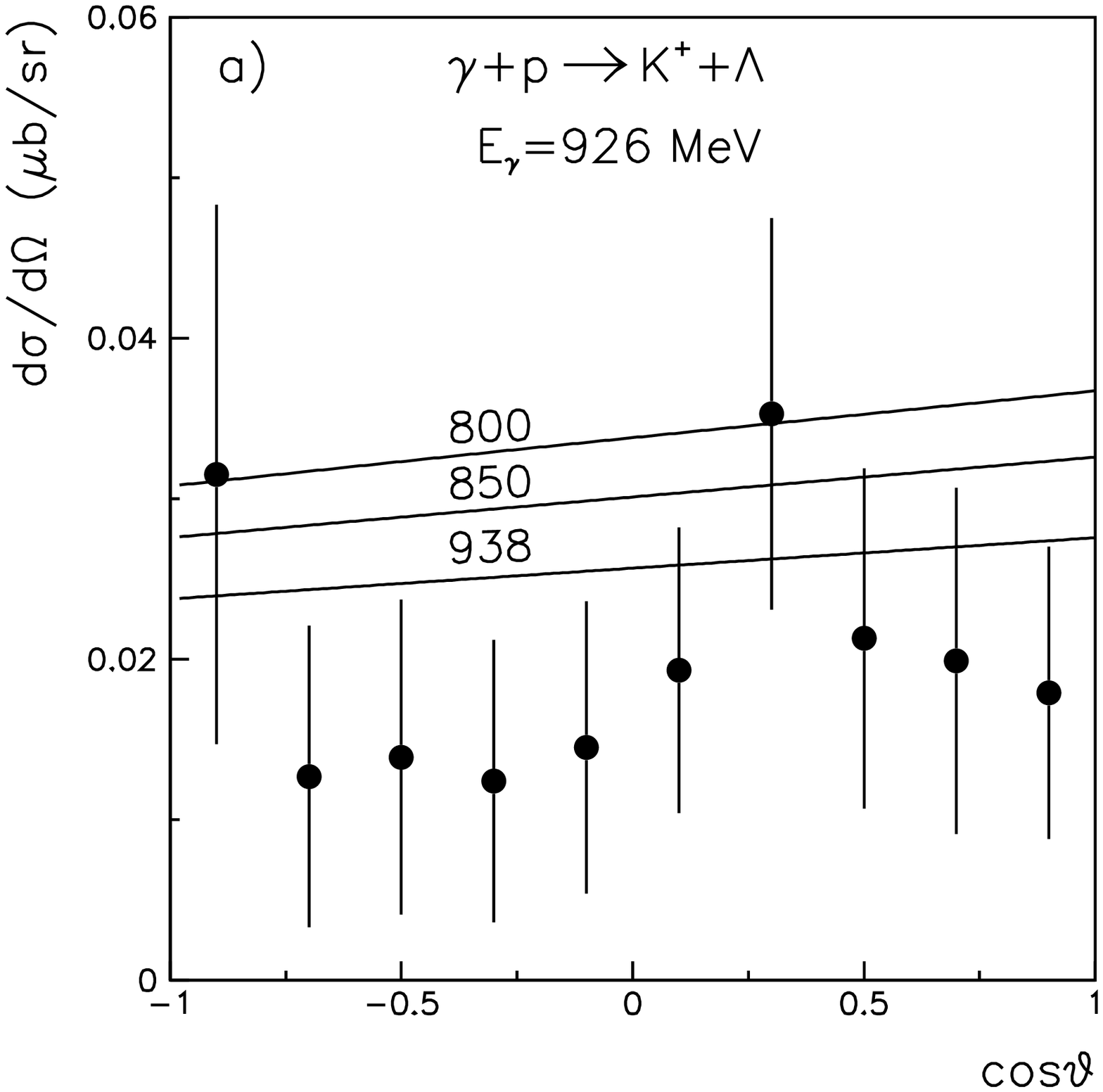,width=8.cm,height=10cm}
\end{minipage}
\begin{minipage}[t]{80mm}\hspace{-10mm}
\psfig{file=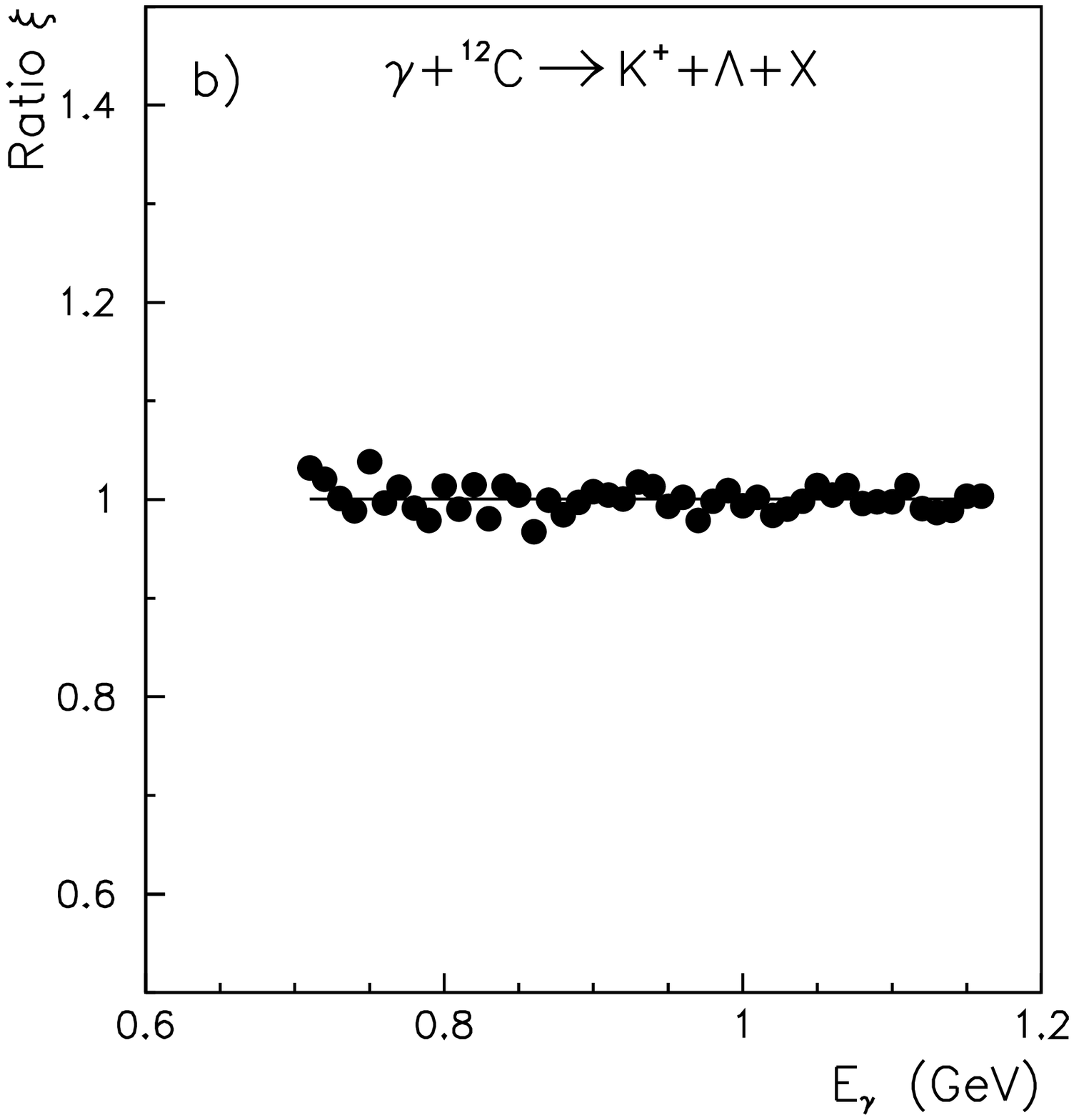,width=8.9cm,height=10cm}
\end{minipage}
\phantom{aa}\vspace{-15mm}
\caption{a) The differential $\gamma{p}{\to}K^+\Lambda$ cross section
at photon energy of 926 MeV calculated for  different  
effective nucleon masses of 800, 850 and 938~MeV as indicated 
in the figure.
The experimental data are from Refs.~\protect\cite{Tran,LB}. 
b) The ratio $\xi$ of the total
$\gamma{C}{\to}K^+\Lambda{X}$ cross section 
calculated with the off-shell elementary amplitude to that
obtained with the on-shell $\gamma{p}{\to}K^+\Lambda$ 
reaction amplitude. The circles show the calculations, while
the solid line indicates the fit for $\xi$=1.0$\pm$0.1}
\label{gaka1d}
\vspace{-2mm}
\end{figure}

Furthermore, the elementary $\gamma{p}{\to}K^+\Lambda$ cross section 
depends on the invariant collision energy $\sqrt{s}$ 
and the off-shellness of the nucleon.
The calculation of $\sqrt{s}$ with the spectral function yield the 
target nucleon energy $E_N$ to be always below the free nucleon mass 
as shown in Fig. 5b). Furthermore, on average the nucleons with the 
highest momenta $q_N$ have the lowest energies $E_N$ (as 
indicated by the full dots in Fig.5b). Thus high momentum components 
go along with large removal energies, which is due to the fact that
two nucleons with opposite high momentum (and low cms momentum) are
correlated. This correlation energy increases even with the nucleon 
momenta and has to be 'paid' extra when trying to extract such a 
nucleon from the nucleus or to use its 4-momentum to produce an 
additional meson. In contrast, within the Thomas-Fermi approximation 
we have $E_N{=}\sqrt{m_N^2+q_N^2}$ and no correlation energy appears.
This explains the sensitivity of the available invariant energy
$\sqrt{s}$ and the production cross section
to the nuclear dispersion relation. 
 
Moreover, we  address the question on modifications
of the  $\gamma{p}{\to}K^+\Lambda$ reaction amplitude
in the nuclear medium.
The nucleon entering the reaction may be off-shell and
its effective mass  no longer be equal to the free nucleon mass.
To illustrate the sensitivity of the amplitude to the 
effective nucleon mass we show in the Fig.~\ref{gaka1d}a) the
differential $\gamma{p}{\to}K^+\Lambda$ cross section
(calculated at photon energy of 926 MeV)  for $m_N$=938,
850 and 800 MeV. The effect is quite noticeable, however, only
for a large mass shift of the nucleon in the medium. Let us recall, 
that  typical single-particles energies for $^{12}C$  are
$\simeq$--34 and $\simeq$--16 MeV and thus we do not expect
a large effect in the production cross section due to 
in-medium modifications of the elementary amplitude. The
actual calculations  confirm these expectations. 

The solid circles in Fig.~\ref{gaka1d}b) show the ratio 
$\xi$ of the total
$\gamma{C}{\to}K^+\Lambda{X}$ cross section 
calculated with the modified elementary amplitude to that
obtained with the  $\gamma{p}{\to}K^+\Lambda$ 
reaction amplitude  in free space. The ratio in Fig.~\ref{gaka1d}b) 
can be  reasonably fitted over the full range of  photon
energies by a constant value $\xi$=1.0$\pm$0.1 
(shown by the solid line).
Thus, in the following calculations we adopt the on-shell 
$\gamma{p}{\to}K^+\Lambda$ reaction amplitude.

\begin{figure}[h]
\phantom{aa}\vspace{-0.6cm}\hspace{12mm}
\psfig{file=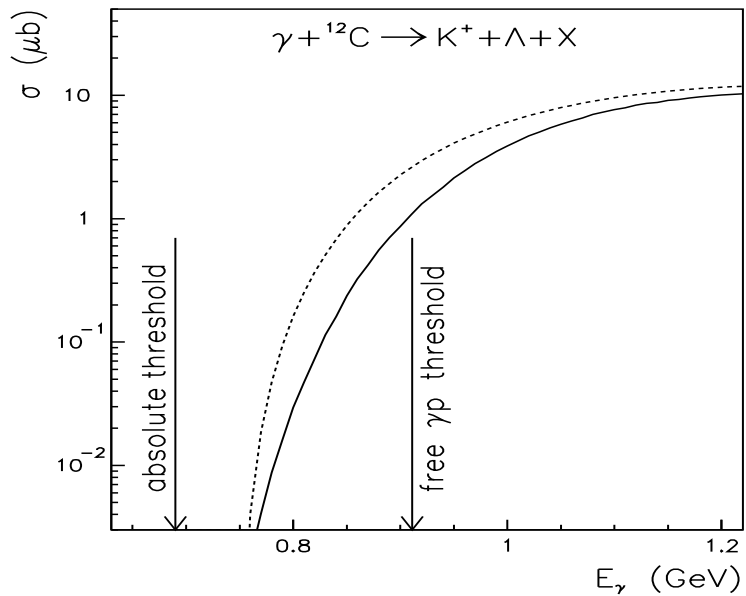,width=12.cm,height=9cm}
\phantom{aa}\vspace{-8mm}
\caption{The $\gamma{C}{\to}K^+\Lambda{X}$ total  cross 
section as a function of the photon energy $E_\gamma$. The  
lines show  calculations with the spectral function (solid)
and Thomas-Fermi model~(\protect\ref{disp1}) (dashed) 
while the arrows indicate
the absolute $\gamma{A}$ and free $\gamma{p}$ reaction threshold.}
\label{gaka2}
\vspace{-2mm}
\end{figure}

The total cross section for the $\gamma{C}{\to}K^+\Lambda{X}$ 
reaction is calculated with  different models 
for $S(q_N, E_N)$ and
shown in Fig.~\ref{gaka2} as a function of the photon
energy. The absolute reaction threshold given by the coherent
production is indicated in  Fig.~\ref{gaka2} by an arrow.
We observe that the result for the total cross section
depends substantially  on the model applied.  At high energies 
$E_\gamma{>}$1.2~GeV both models roughly give the same total cross 
section. At $E_\gamma{\simeq}800$~MeV the difference between the
calculations with the spectral function and the Thomas-Fermi approach
amounts to a factor of about 6, which results from the
different dispersion relations applied to the target nucleon.
This can easily be understood by inspection of 
Fig.~\ref{gaka8}b).

One of the direct ways to extract the dispersion relation
can be understood in terms of the four momentum squared $Q_N^2$ 
of the target nucleon. Within the Thomas-Fermi model it is 
given as
\begin{equation}
Q_N^2 \simeq (m_N + U)^2,
\end{equation}
where we imply $q_N{\ll}m_N$. With the potential
given by Eq.~(\ref{pwelke}) one can estimate 
$Q_N^2{\simeq}$ 0.8~GeV$^2$. Moreover, within the Thomas-Fermi 
approximation $Q_N^2$ is always positive and does not depend
on the photon energy.

\begin{figure}[h]
\phantom{aa}\vspace{-8mm}
\begin{minipage}[t]{80mm}
\psfig{file=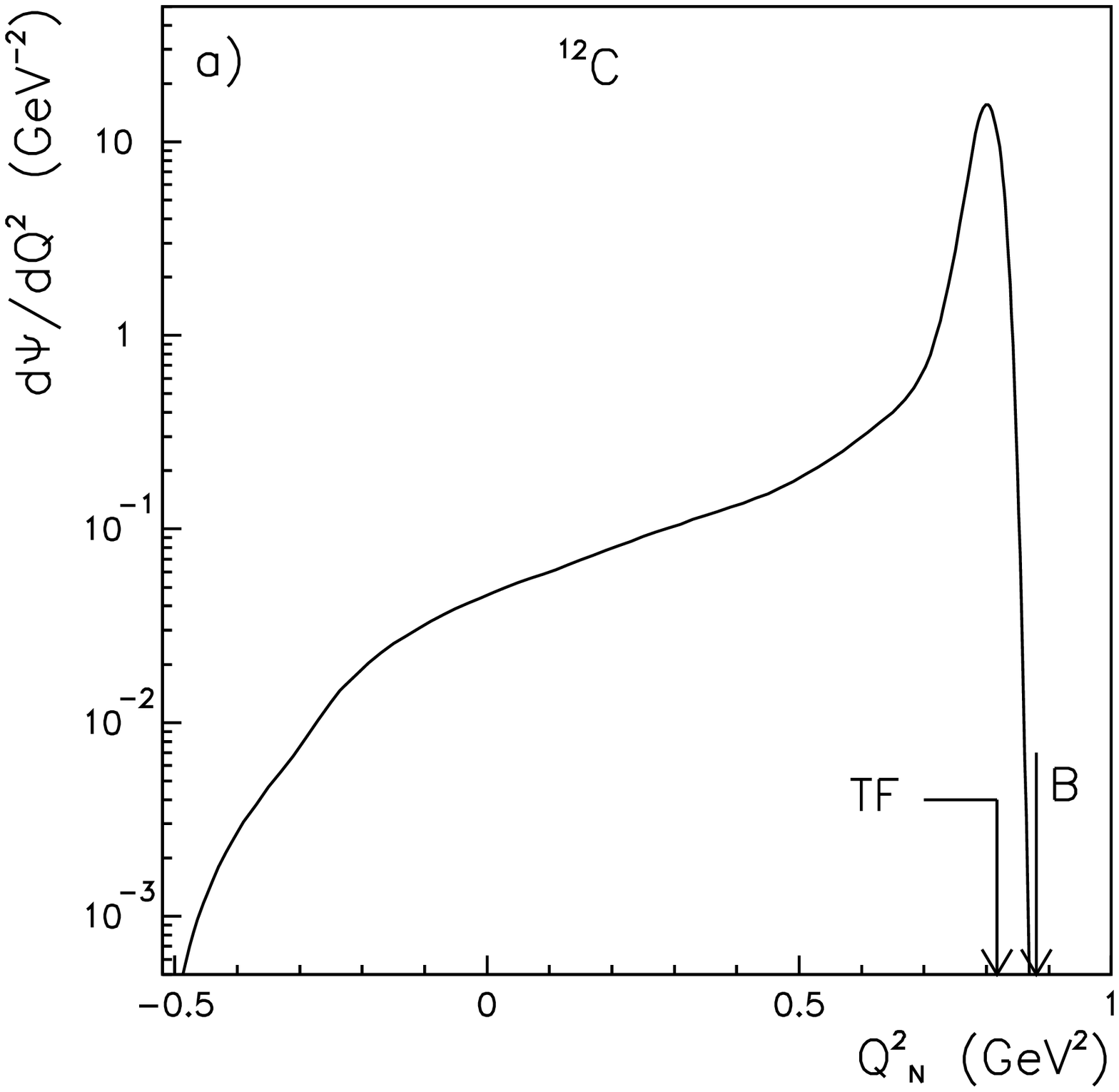,width=8.3cm,height=10cm}
\end{minipage}
\phantom{aa}\hspace{-15mm}
\begin{minipage}[t]{80mm}
\psfig{file=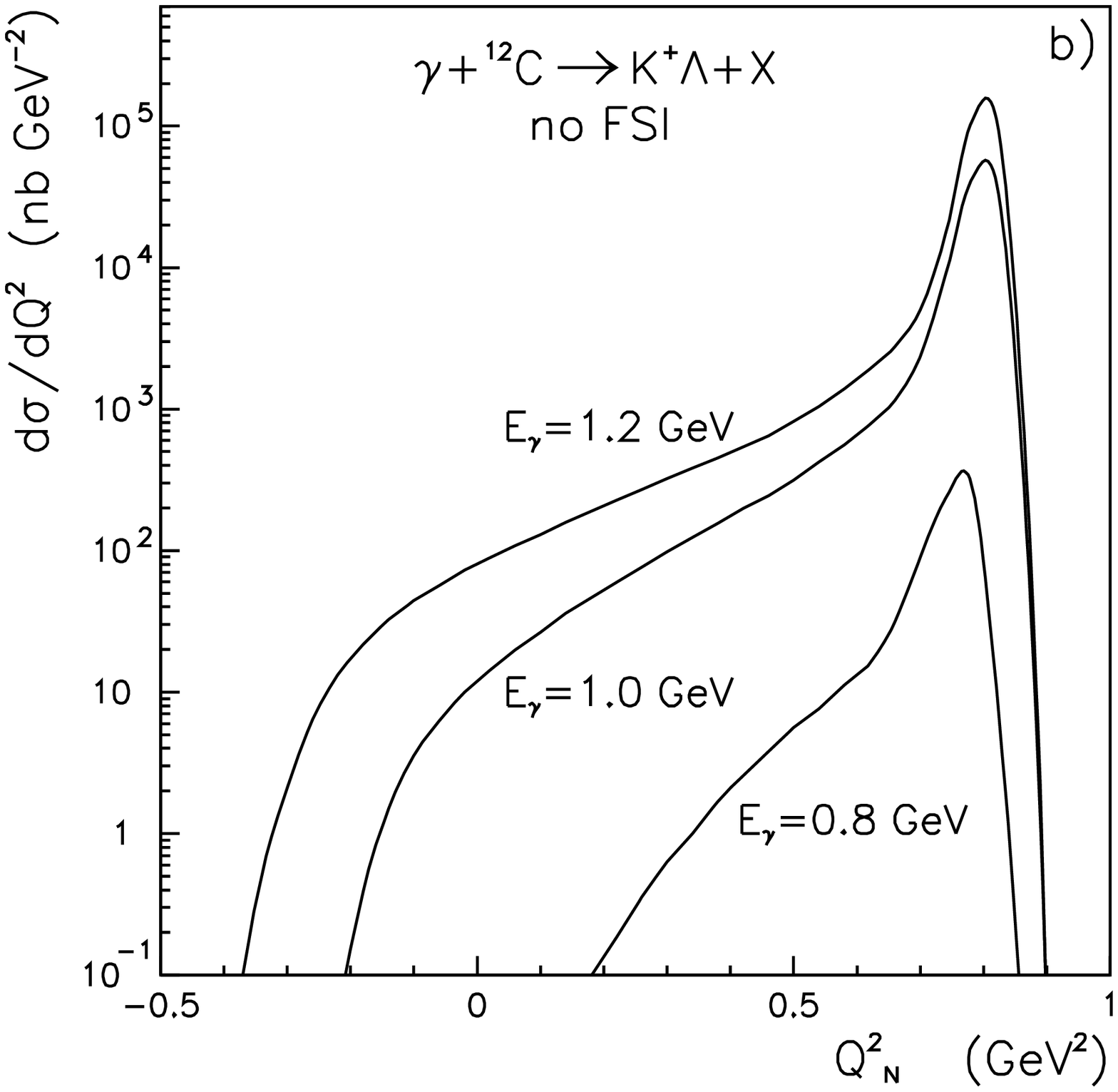,width=8.3cm,height=10cm}
\end{minipage}
\phantom{aa}\vspace{-15mm}
\caption{a) The $Q^2_N$ spectrum $d\Psi/dQ^2_N$ 
given by the spectral function for
$^{12}C$. The arrows denote the Thomas-Fermi 
limit~(\protect\ref{disp1}) (TH) and 
the squared nucleon mass $B$. b) The differential $Q^2_N$
spectra calculated for the $\gamma{C}{\to}K^+\Lambda{X}$ reaction
at different photon energies $E_\gamma$.}
\label{gaka12}
\vspace{-6mm}
\end{figure}

Within the spectral function approach the quantity $Q_N^2$ is
distributed as 
\begin{equation}
\frac{d\Psi}{dQ^2_N} = \oint d^3q_N \, dE_N \,S(q_N, E_N)
\label{count}
\end{equation}
where the integration is performed over the contour 
$Q^2{=}E_N^2{-}q_N^2$. The function~(\ref{count}) is
shown in Fig.~\ref{gaka12}a). The negative part of the
$Q_N^2$ spectrum corresponds to  high nucleon momenta $q_N$ and
small energies $E_N$. In Fig.~\ref{gaka12}a) the arrow denoted
by $B$ shows the squared bare mass of the nucleon, while
the arrow denoted by $TF$ indicates the  value for $Q^2_N$ as 
given by the Thomas-Fermi approximation~(\ref{disp1}).
It is seen that the $Q^2$ distribution extends well
below $B$.

It is clear that at photon energies above the reaction 
threshold in free space the total $Q_N^2$ spectrum, as given by 
the spectral function, is available for  $K^+\Lambda$ 
production. Fig.~\ref{gaka12}b) shows the $Q^2_N$ spectra 
calculated for the $\gamma{C}{\to}K^+\Lambda{X}$ reaction at 
$E_\gamma$=0.8, 1.0 and 1.2~GeV.

The  $Q^2_N$ spectrum can be obtained experimentally  
from  the four-momenta of the
produced $K^+$-mesons and hyperons as
\begin{equation}
Q^2_N = E_N^2-q_N^2 = (E_K +E_\Lambda - E_\gamma)^2 -
({\bf p}_K + {\bf p}_\Lambda -{\bf k}_\gamma )^2
\label{recor}
\end{equation}
Note, that  relation~(\ref{recor}) is similar to 
the experimental analysis for the reconstruction of the 
spectral function $S(E_N,q_N)$ in  electron-nucleus
scattering~\cite{Frullani}. We also note, that
the $Q^2_N$ spectra shown in Fig.~\ref{gaka12}b)
might be substantially distorted due to the final state 
interaction (FSI) of both the  $K^+$-mesons and the $\Lambda$-hyperons.
Similar problems also appear in  $eA{\to}e^{\prime}pX$
studies.

\begin{figure}[h]
\phantom{aa}\vspace{-8mm}
\begin{minipage}[t]{80mm}
\psfig{file=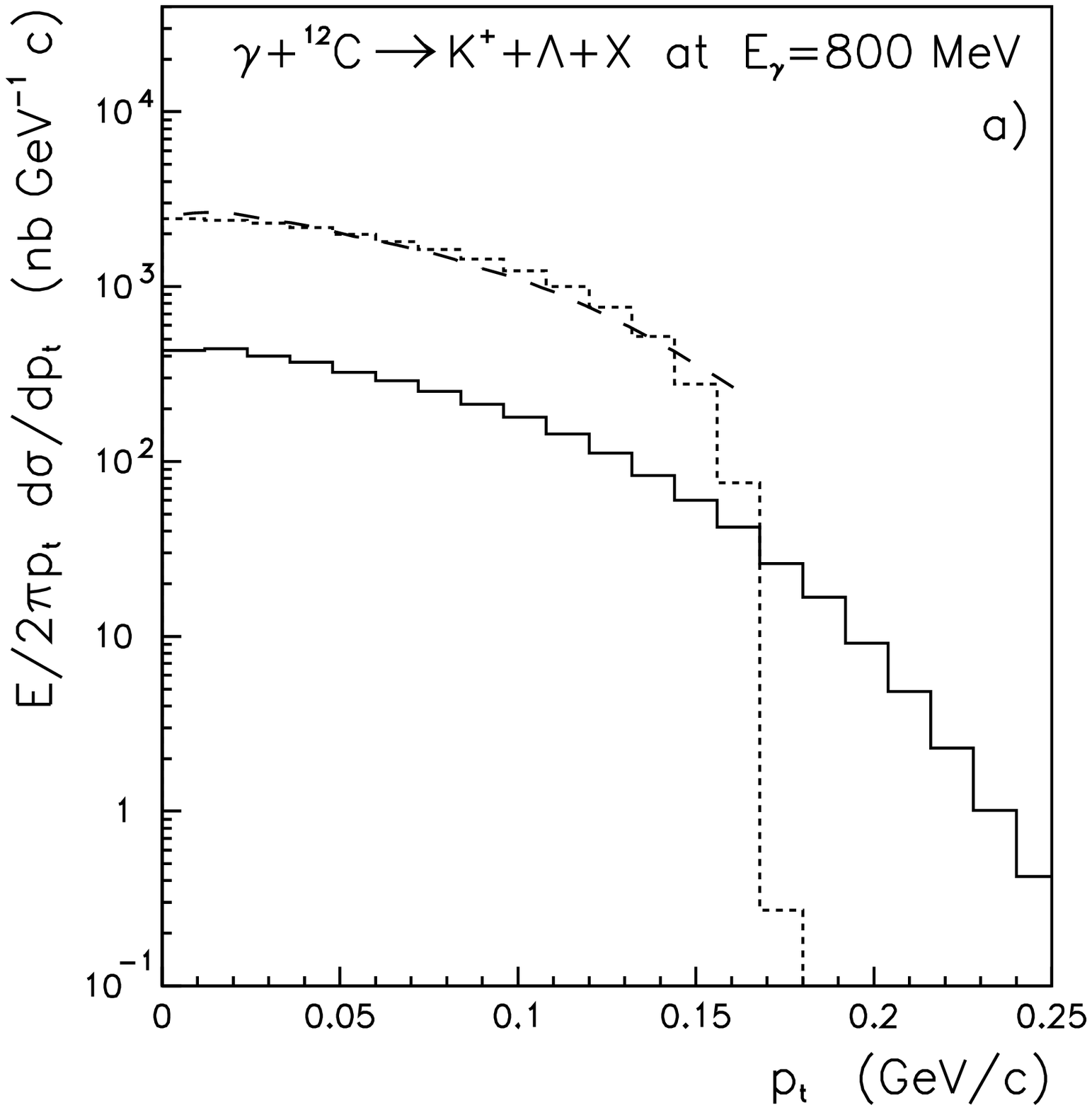,width=8.3cm,height=10cm}
\end{minipage}
\phantom{aa}\hspace{-15mm}
\begin{minipage}[t]{80mm}
\psfig{file=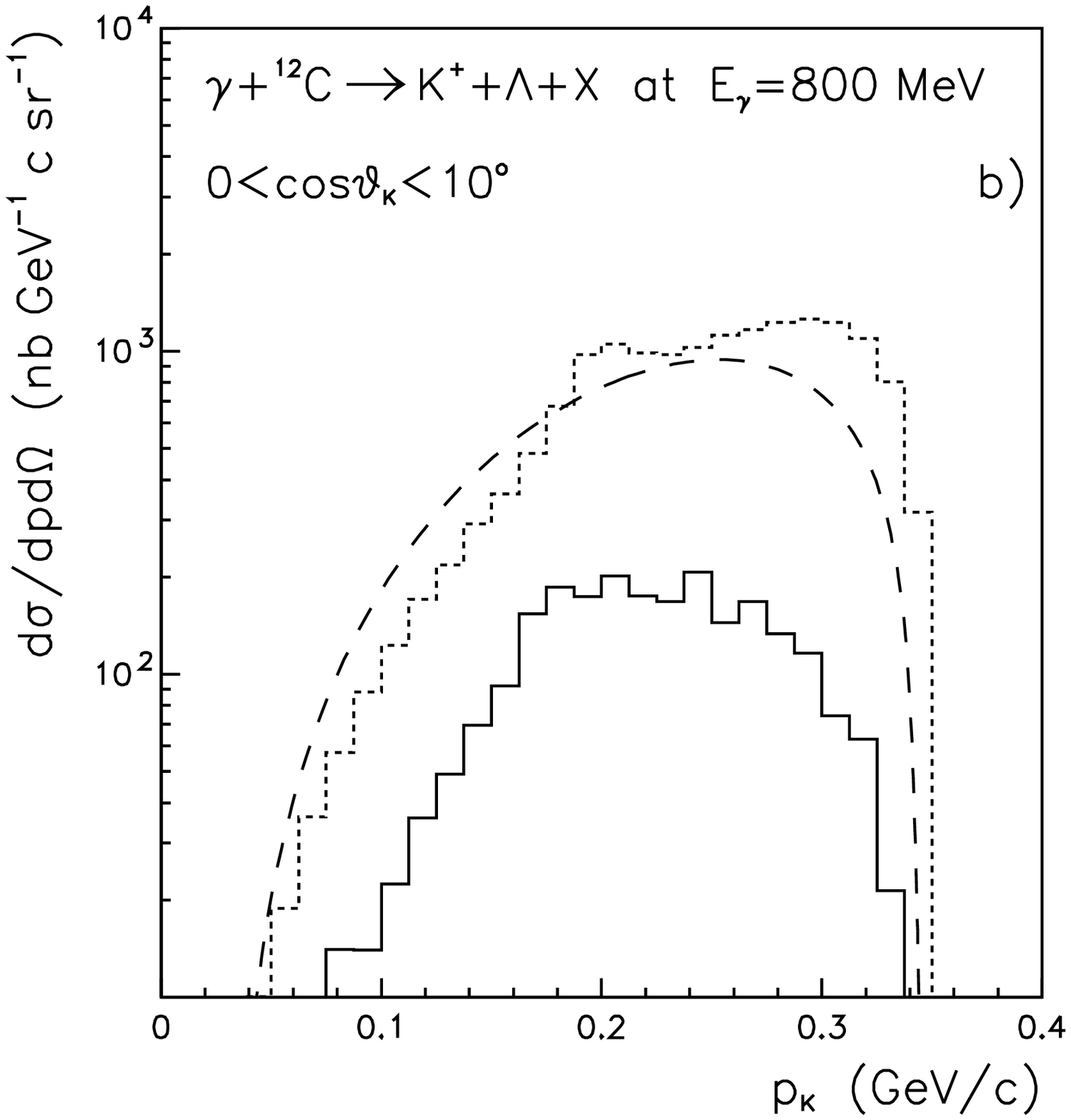,width=8.3cm,height=10cm}
\end{minipage}
\phantom{aa}\vspace{-15mm}
\caption{The transverse (a) and total momentum spectrum (b) at
$0{<}cos\theta{<}10^0$ of $K^+$-mesons from the 
$\gamma{C}{\to}K^+\Lambda{X}$ reaction at  a photon energy
of 800 MeV. The histograms show the calculations with the 
spectral function (solid) and Thomas-Fermi model while
the long-dashed line is the calculation with the spectral function
multiplied by a factor of 6.}
\label{gaka6}
\vspace{-12mm}
\end{figure}

Another sensitive observable is  the transverse momentum
($p_t$) spectrum of the produced kaons. In the elementary 
interaction between the photon and the target nucleon the 
transverse component of the total initial momentum is given by 
the nucleon momentum $q_N$, which is different in 
the Thomas-Fermi and the spectral function approximation. 
Due to momentum conservation the $p_t$ distribution of the
produced $K^+$-mesons reflects the transverse part
of the $q_N$ distribution.

The $p_t$ spectrum for $K^+$-mesons from the 
$\gamma{C}{\to}K^+\Lambda{X}$ 
reaction at $E_\gamma$=800~MeV is calculated within the two 
models and shown in Fig.~\ref{gaka6}a). The solid histogram shows 
the calculations with the spectral function, the dashed one
that obtained with the Thomas-Fermi approach. The results indicate 
that the absolute magnitudes of the differential cross sections
from the two models are substantially different for $p_t{<}$170~MeV/c
in line with the result shown in Fig.~\ref{gaka2}.
The solid line in Fig.~\ref{gaka6}a) shows the calculations
with the  spectral function multiplied by a factor of 6
in order to account for the absolute difference in  the total 
production cross section from the models.
The  shape of the $p_t$ spectra from the two 
models differs at large $p_t{>}$170~MeV/c since  in the 
Thomas-Fermi approximation the nucleon  momenta $q_N$ are limited 
to values below $q_F$.

Fig.~\ref{gaka6}b) shows the momentum spectrum of
$K^+$-mesons from the $\gamma{C}{\to}K^+\Lambda{X}$ 
reaction at $E_\gamma$=800~MeV integrated over the
laboratory angle $0{<}cos\theta{<}10^0$. Again we find a 
substantial  difference between the calculations with 
the spectral function (solid histogram) and Thomas-Fermi 
approximation (dashed histogram) in
the absolute value of the production cross section.

To compare the shapes of the momentum spectra we have
renormalized the calculations with the spectral function
and show this result in Fig.~\ref{gaka6}b) by the long-dashed line.
Both models provide  different shapes of the
momentum spectrum. The calculations with the  Thomas-Fermi 
model indicate a detectable enhancement at large $K^+$-meson 
momenta. A measurement of the $K^+$ spectra from 
$\gamma{C}{\to}K^+\Lambda{X}$ reaction at $E_\gamma$=800~MeV
and at forward angles is thus sensitive to the dispersion relation.

\section{Contribution from secondary processes}
In principle, the $K^+\Lambda$ production in ${\gamma}A$
interactions can also proceed through a two-step process with
an intermediate pion from the ${\gamma}N{\to}\pi{N}$ reaction followed by 
the strangeness production channel
${\pi}N{\to}K^+\Lambda$. The cross sections for $K^+$-meson 
production from the secondary pion induced reactions is 
calculated as~\cite{Sibirtsev3,Cassing4,Buescher}
\begin{eqnarray}
E_K \frac {d^3 \sigma_{\gamma A\to K \Lambda X}} {d^3 p_K} =
\int\limits_{\Omega}  d^3 q_N \ dE_N
\frac { d^3 p_{\pi }^\prime } {E_{\pi}^\prime}
\ S(q_N, E_N) \ A_{eff}(\sigma_{\pi N}, \sigma_{KN})
\nonumber \\
\times 
\frac { 1}{\sigma_{tot} }  E_K^{\prime\prime}
\frac {d^3 \sigma_{\pi N\to K \Lambda}(\sqrt{s})}
{d^3p_M^{\prime\prime} }
\  E_{\pi}^{\prime}
\frac { d^3 {\sigma}_{\gamma N\to \pi N}(\sqrt{s^{\prime}}) }
{d^3 p_{\pi }^{\prime} },
\label{second}
\end{eqnarray}
where the  double prime indices denote the system of the
intermediate pion and a target nucleon, while the single prime 
indices are those for the photon and the first target nucleon. 
Moreover,
$ E_{\pi} d^3{\sigma}_{\gamma N\to \pi N}/d^3p_{\pi}$
stands for the $\pi$-meson differential production 
cross section, which is calculated
in line with Eq.~(\ref{direct}) assuming on-shell pions in the 
nuclear medium,  while ${\sigma}_{tot}$ is the total
$\pi{N}$ cross section.

In  Eq.~(\ref{second}) the index $\Omega$ at the integral 
stands for the Pauli
blocking of the final states from the ${\gamma}N{\to}\pi{N}$
reaction. Furthermore, the factor $A_{eff}$ in Eq.~(\ref{second})
accounts for the effective primary collision number, pion 
interaction in the nuclear interior as well as for the in-medium 
distortion of the produced $K^+$-meson. It is calculated by
an integration over the path of the produced $\pi$-mesons and
final kaons as proposed in Ref.~\cite{Vercellin,Effenberger}.

The elementary  ${\gamma}N{\to}\pi{N}$ reaction  
is calculated in line with the  results from the Virginia 
WI98K partial wave analysis~\cite{Arndt}. Note, 
that three reaction channels: 
${\gamma}p{\to}\pi^+n$, ${\gamma}p{\to}\pi^0p$ and 
${\gamma}n{\to}\pi^0n$ enter
the two-step process calculations of $K^+\Lambda$ photoproduction.

\begin{figure}[h]
\phantom{aa}\vspace{-8mm}
\begin{minipage}[t]{80mm}
\psfig{file=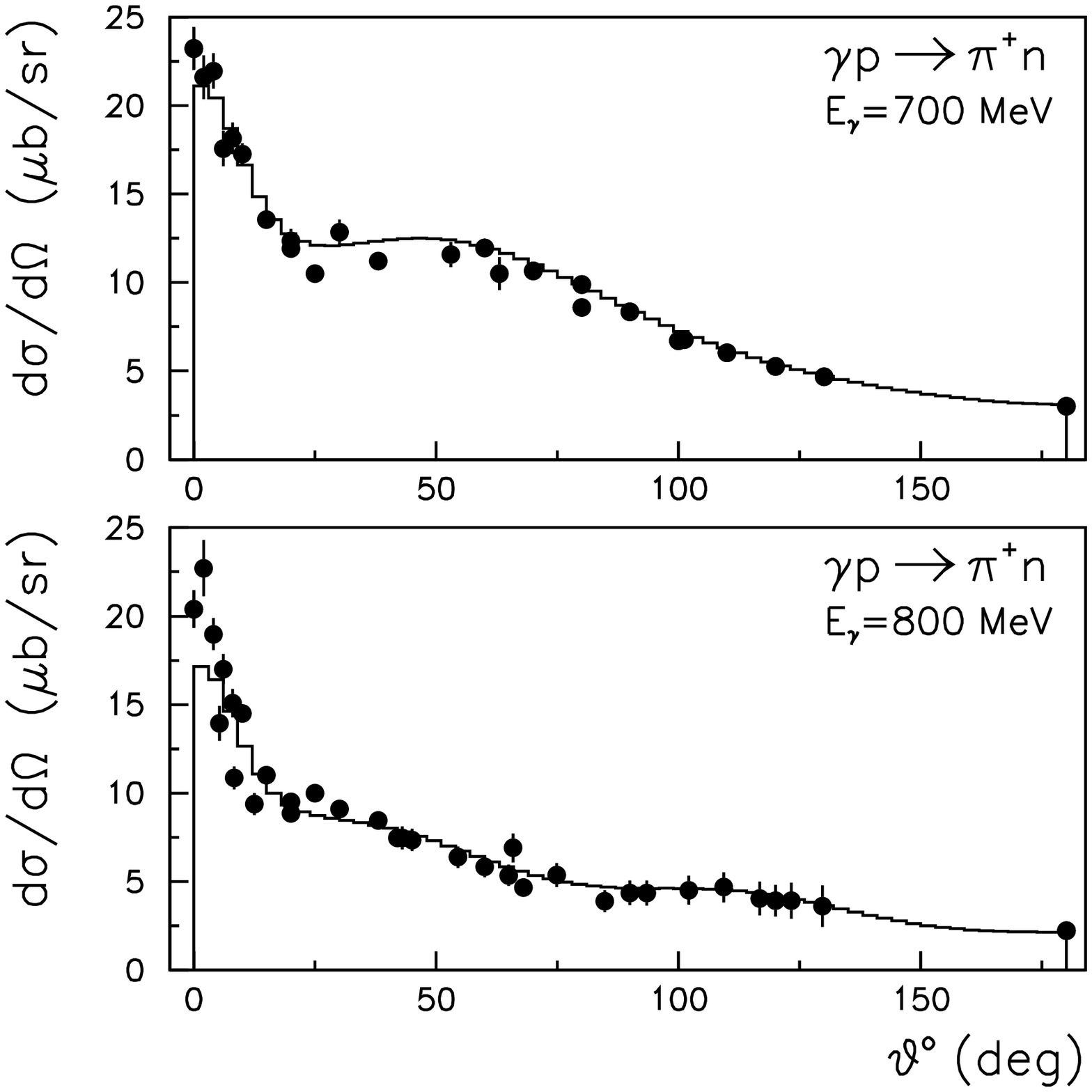,width=8.3cm,height=11cm}
\end{minipage}
\phantom{aa}\hspace{-15mm}
\begin{minipage}[t]{80mm}
\psfig{file=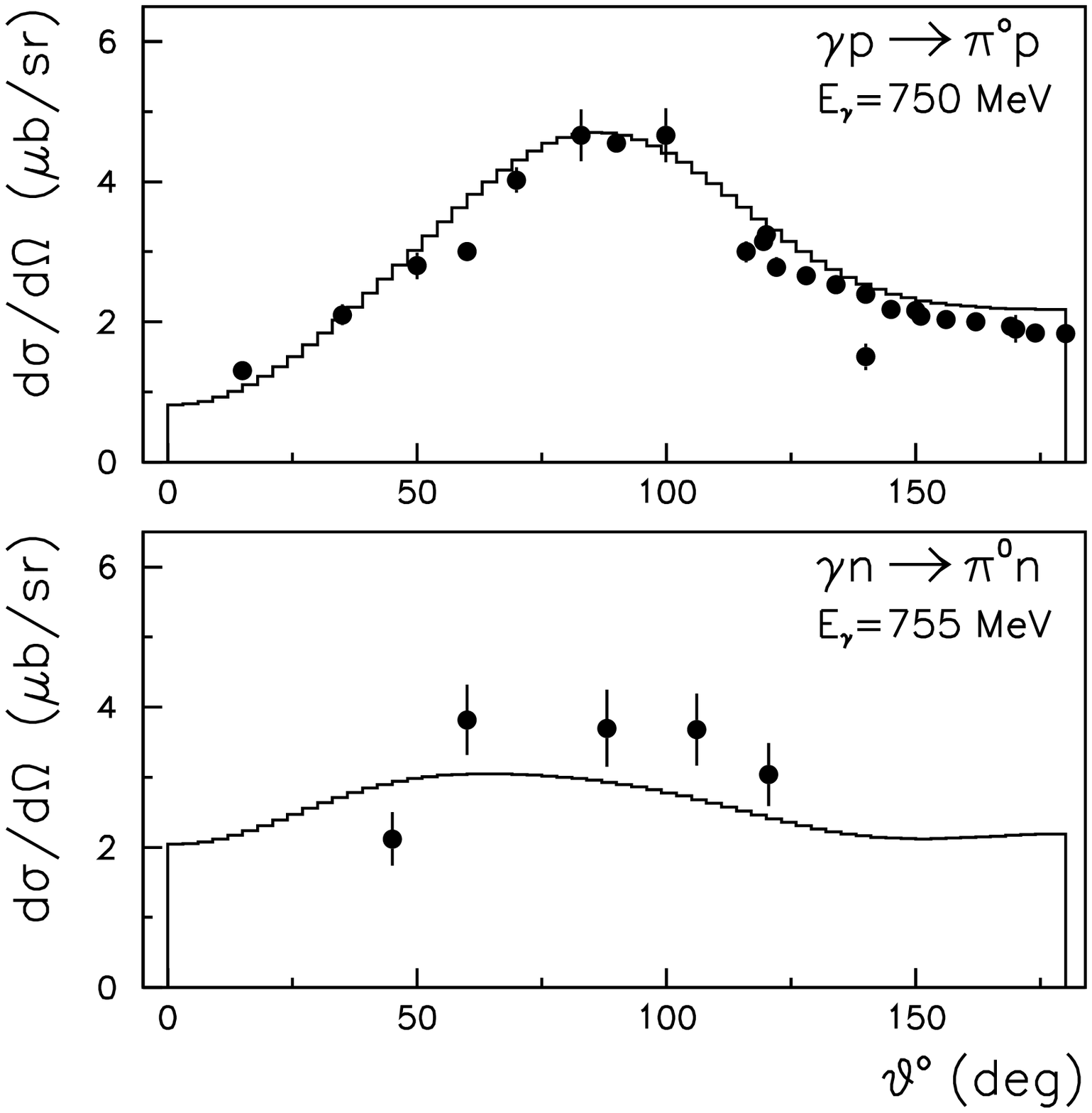,width=8.3cm,height=11cm}
\end{minipage}
\phantom{aa}\vspace{-15mm}
\caption{The differential cross sections for different channels
of the ${\gamma}N{\to}\pi{N}$ reaction at several photon energies.
The circles show the experimental data~\protect\cite{LB,Arndt1} while the
histograms indicate our simulations based on the Virginia 
WI98K partial wave analysis~\protect\cite{Arndt}.}
\label{gaka1bc}
\vspace{-2mm}
\end{figure}

Fig.~\ref{gaka1bc} shows the experimental data~\cite{LB,Arndt1} 
on the differential cross sections for various
reaction channels at several photon energies $E_\gamma$
in comparison with the simulation code. Since the  WI98K
partial wave solution describes the data up to  $E_\gamma$=2~GeV,
we expect to be accurate  within
this energy range. The cross sections for the 
$\pi^0p{\to}K^+\Lambda$ and $\pi^+n{\to}K^+\Lambda$ reactions 
are taken from the resonance model calculation of
Tsushima et al.~\cite{Tsushima1,Tsushima2}. As  shown in 
Refs.~\cite{Tsushima1,Tsushima2} the resonance model reproduces 
the data well up to  $K^+\Lambda$ invariant masses of $\simeq$2~GeV.

\begin{figure}[h]
\phantom{aa}\vspace{-8mm}\hspace{12mm}
\psfig{file=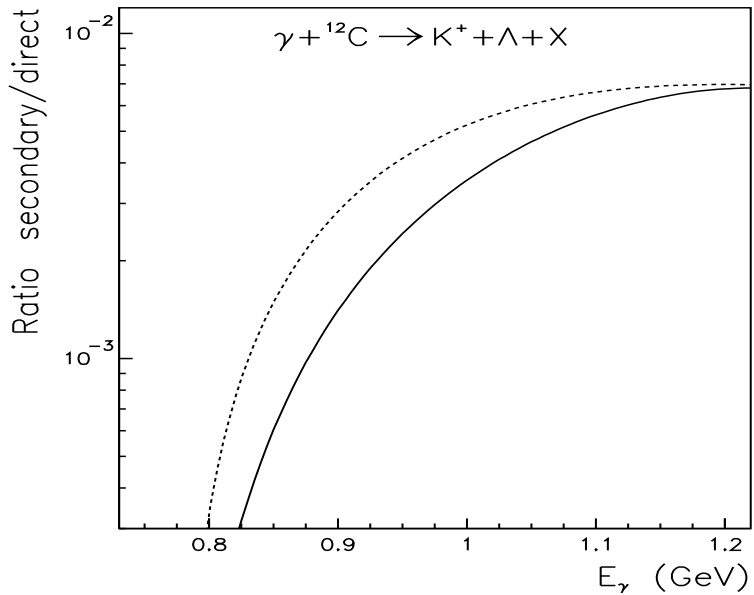,width=12.cm,height=9cm}
\phantom{aa}\vspace{-8mm}
\caption{The ratio of the contribution to the
$\gamma{C}{\to}K^+\Lambda{X}$ reaction cross 
section from two-step to direct production process
as a function of the photon energy $E_\gamma$. The dashed
line shows the  calculations  within the Thomas-Fermi 
model~(\protect\ref{disp1}) while the
solid line gives the result of the spectral function approach.}
\label{gaka2a}
\vspace{-2mm}
\end{figure}

Here we do not account for  multi-pion production in
$\gamma{N}$ reactions, which might slightly enhance the contribution
from the secondary processes. Since the $\pi$-meson spectra
from multi-pion production at $E_\gamma{\le}$1~GeV are 
soft~\cite{Effenberger2} and have their main strengh
essentially below the $\pi{N}{\to}K^+\Lambda$ threshold, we 
consider our estimate for the contribution from secondary processes 
as a lower limit. 

The dashed line in Fig.~\ref{gaka2a} shows the ratio of 
the $K^+\Lambda$ production cross section from secondary processes 
to the contribution from the direct production 
calculated within the Thomas-Fermi  approximation. At 
$E_\gamma{<}1.2$~GeV the contribution from the two-step reaction
mechanism amounts to less than  1\%. The solid line 
in Fig.~\ref{gaka2a} shows the calculations with the spectral 
function for comparison.
We thus conclude that the contribution from the two-step process
to the $\gamma{C}{\to}K^+\Lambda{X}$ reaction is almost
negligible at photon energies below 1.2~GeV. 

In our study on subtreshold meson production from $pA$
collisions~\cite{Sibirtsev3} it was found that the contribution 
from secondary pion induced reactions dominates the direct
production mechanism. This finding is different from our 
present result obtained for the $\gamma{A}$ reactions and
needs further explanation. 

\begin{figure}[h]
\phantom{aa}\vspace{-8mm}
\begin{minipage}[t]{80mm}
\psfig{file=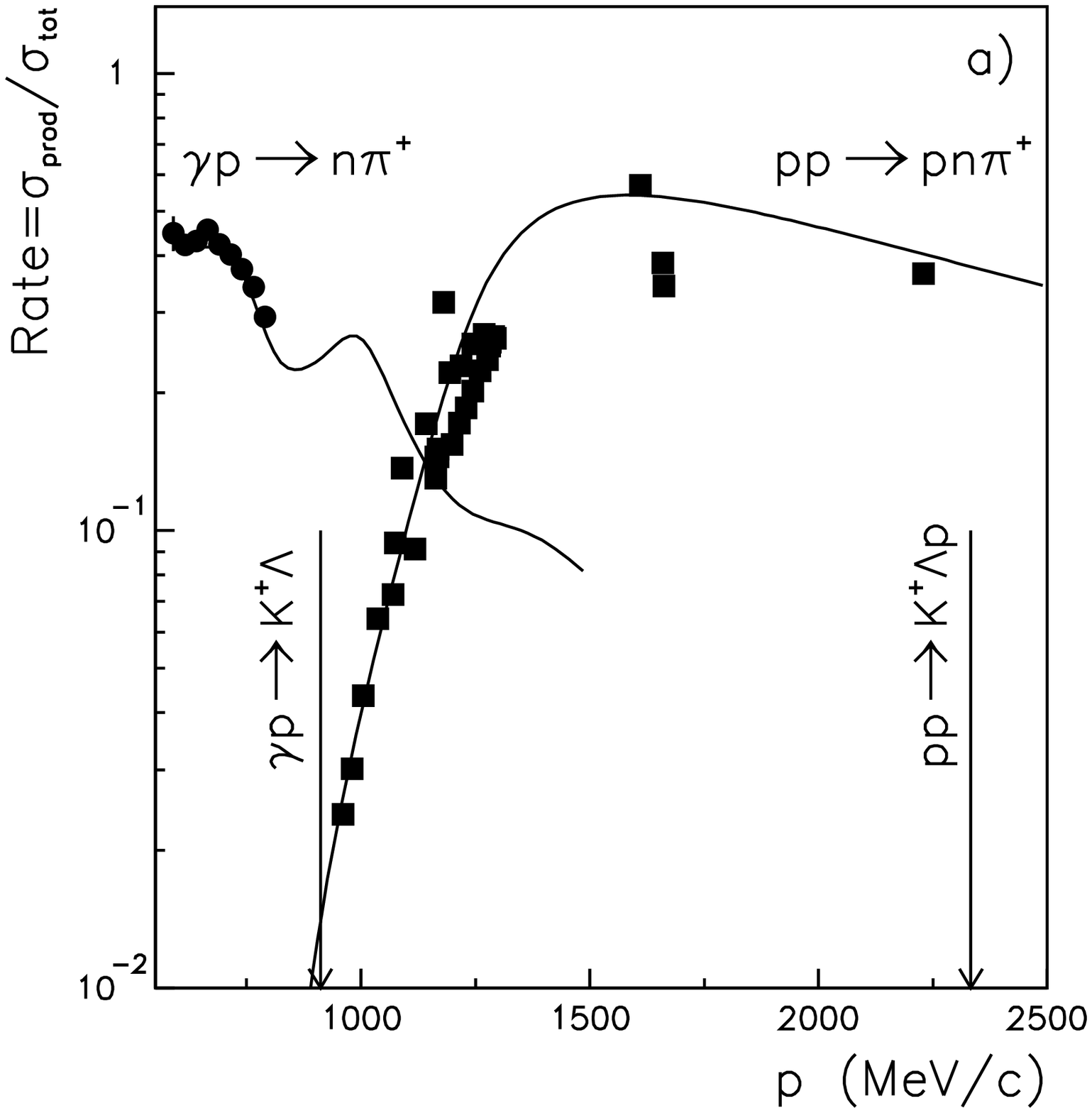,width=8.3cm,height=10cm}
\end{minipage}
\phantom{aa}\hspace{-15mm}
\begin{minipage}[t]{80mm}
\psfig{file=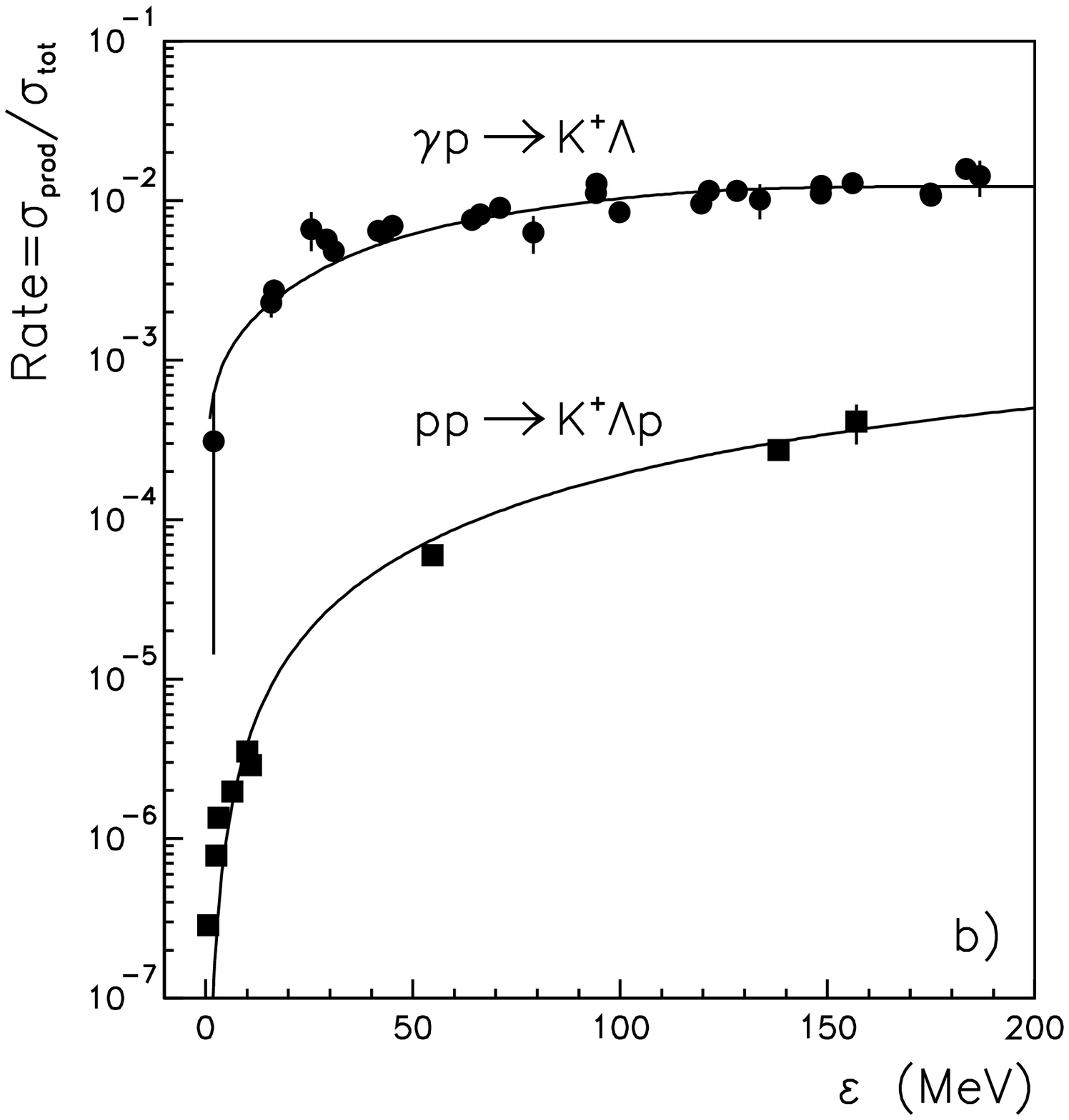,width=8.3cm,height=10cm}
\end{minipage}
\phantom{aa}\vspace{-15mm}
\caption{a) The rates of the $\gamma{p}{\to}n\pi^+$ and
$pp{\to}pn\pi^+$ reactions as a function of the projectile 
momentum $p$.
The arrow indicates the thresholds for the direct  
${\gamma}p{\to}K^+\Lambda$ and $pp{\to}K^+\Lambda{p}$ production.
b) The rates of the direct ${\gamma}p{\to}K^+\Lambda$ and 
$pp{\to}K^+\Lambda{p}$ production as a function of the excess
energy $\epsilon$. The data are 
from Ref.~\cite{LB} while the lines show our
calculations.}
\label{gaka2bc}
\vspace{-2mm}
\end{figure}

Fig.~\ref{gaka2bc}a) shows the rate of the pion production
in $\gamma{p}$ and $pp$ interactions as a function of the projectile
momentum. This rate is defined as a ratio of the $\pi$-meson
production cross section to the total reaction cross section. 
Let us now consider the role of the secondary reaction
mechanism at energies below the direct ${\gamma}p{\to}K^+\Lambda$
and $pp{\to}K^+\Lambda{p}$ production thresholds, which
are indicated in  Fig.~\ref{gaka2bc}a) by the arrows. 
It is clear that near  the direct production 
threshold the $\pi$-meson production rate is almost the same for the
$\gamma$ and $p$ induced reaction. We thus conclude that the strength 
of the secondary process at a given projectile momentum is also 
almost the same for the photon and proton induced reactions on nuclei.

Fig.~\ref{gaka2bc}b) shows the rates of  direct $K^+\Lambda$ 
production from $\gamma{p}$ and $pp$ reactions as a function of 
the excess energy $\epsilon$. Here $\epsilon$ is given as
the difference between the invariant collision energy and the 
total mass of the final particle. Both the experimental data 
and our calculations indicate that the rate of the  
$\gamma{p}{\to}K^+\Lambda$ reaction is substantially ( by about two
orders of magnitude) larger than the $pp{\to}K^+\Lambda{p}$
rate at energies close to the threshold in free space.
Thus the direct $K^+\Lambda$ production dominates in
photon induced reactions on nuclei and the cross section is
thus directly sensitive to the spectral function of the nucleon 
in the target nucleus. 

This is an important finding indicating the advantage of 
$\gamma{A}$ reactions in styding the production at energies
below the reaction threshold on a free nucleon. It is presently an open
question if the production of  
heavy mesons such as $\rho$, $\omega$ and $\phi$ show the same tendency.
We also refer to Ref.~\cite{Effenberger2}, where a  
dominance of the direct production mechanism was found for 
subthreshold $\eta$-meson production in $\gamma{A}$ reactions.

\section{Conclusions}
We have studied  $K^+$-meson production in $\gamma{C}$ interactions
at energies below the free $\gamma{p}{\to}K^+\Lambda{X}$
reaction threshold. The Thomas-Fermi and spectral function
approximation were used in the calculation of the  production 
process. It was found that these two models result in quite
different total and differential cross sections for kaon
photoproduction.

The difference between the two models for the target nucleon 
distribution can 
be detected in kaon photoproduction not  only 
in the absolute scale, but also in  the different shape of the 
spectra. The shapes of the transverse momentum spectra
calculated with the two models seem to be identical 
at low transverse momenta $p_t{<}$170~MeV/c, but
differ substantially at larger $p_t$. The large $p_t$ region is 
sensitive to the details of the spectral function 
although the contribution from  $p_t{>}$170~MeV/c to the total  
$\gamma{C}{\to}K^+\Lambda{X}$ cross section is  very small. 

Moreover, we have found that the evaluation of the nuclear spectral function
from the  $\gamma{A}{\to}K^+\Lambda{X}$ data by  the
$K^+$-meson transverse spectra or the reconstruction of the 
squared four momentum $Q_N^2$ (see Fig.~\ref{gaka12}b))
of the bound nucleon might be possible if the final state interactions
of the produced particles are small. As can  be seen from Fig.~\ref{gaka2}
reactions slightly above this threshold are favoured experimentally
due to the higher cross section.

We have, furthermore, found that the contribution from secondary process
$\gamma{N}{\to}\pi{N}$, $\pi{N}{\to}K^+\Lambda$ to the
kaon photoproduction is almost negligible for 
$E_\gamma{\le}$1.2~GeV. 

\vspace{4mm}
The authors like to acknowledge discussions with T. Feuster and
G. Penner and valuable comments and suggestions from A. Gal
for the final form of this work.

\end{document}